%% file: BXsec.tex
\documentclass[preprint,12pt,3p]{elsarticle}
\pdfoutput=1

\usepackage{definitions} 
\usepackage{atlasphysics} 
\usepackage[hyperindex,breaklinks]{hyperref}
\usepackage{preprintcover}  
\usepackage{amssymb}
\usepackage{mathrsfs}
\usepackage{subfigure}
\usepackage{ifthen}
\newboolean{supp}
\setboolean{supp}{false}
\usepackage{lineno}

\biboptions{sort&compress}

\journal{Nuclear Physics B}

\PreprintCoverPaperTitle{Measurement of the $b$-hadron production cross section using decays to $\dsmu X$ final states in $pp$ collisions at $\sqrt{s}=7$ 	\TeV{} with the ATLAS detector} 

\PreprintIdNumber{CERN-PH-EP-2012-121}  

\PreprintCoverAbstract{
\input{abstr}
}  
\PreprintJournalName{Nuclear Physics B}

\begin{document}

\begin{frontmatter}

\title{Measurement of the $b$-hadron production cross section using decays to $\dsmu X$ final states in $pp$ collisions at $\sqrt{s}=7$ 	\TeV{} with the ATLAS detector}

\author{The ATLAS Collaboration}

\address{}

\begin{abstract}

\input{abstr}
\end{abstract}


\end{frontmatter}

\input{intro}
\input{detector}
\input{outline}
\input{montecarlo}
\input{selection}

\input{samplecomp}

\input{efficiency}

\input{systematics}
\input{unfolding}

\input{sigma}

\input{discussion}
\input{conclusion}
\input{acknowledgements}

\bibliographystyle{model1-num-names}

\clearpage
\input{atlas_authlist}

\end{document}

%% file: abstr.tex
The $b$-hadron production cross section is measured with the ATLAS detector in $pp$ collisions at $\sqrt{s}=7\TeV$, using 3.3\,\ipb{} of integrated luminosity, collected during the 2010 LHC run.
The $b$-hadrons are selected by partially reconstructing $\dsmu X$ final states. Differential cross sections are measured as functions of the transverse momentum and pseudorapidity. The measured production cross section for a $b$-hadron with $\pT>9\GeV$ and $|\eta|<2.5$ is $32.7 \pm 0.8\stat  ^{+4.5}_{-6.8}\syst   \,\,\mu\mbox{b}$, higher than the next-to-leading-order QCD predictions but consistent within the experimental and theoretical uncertainties.

%% file: intro.tex
\section{Introduction}
\label{sec:intro}

The production of heavy quarks at hadron colliders provides a challenging opportunity to test the validity of quantum chromodynamics (QCD) predictions and calculations. The $b$-hadron production cross section has been predicted with next-to-leading-order (NLO) accuracy for more than twenty years~\cite{theory1,theory2}.

Several measurements were performed with proton-antiproton collisions by the UA1 experiment at the Sp$\bar{\mathrm{p}}$S collider (CERN) at a centre-of-mass energy of $\sqrt{s}=630\GeV{}$~\cite{ua1,ua2}, and by the CDF and D0 experiments at the Tevatron collider (Fermilab) 
at $\sqrt{s}=630\GeV{}$, $1.8\TeV{}$ and $1.96\TeV{}$~\cite{cdf0,cdf1,cdf2,cdf3,d01,d02,d03,cdf4,cdf5,cdf6}.
These measurements made a significant contribution to the understanding of heavy-quark production in hadronic collisions~\cite{cacciari}, but the theoretical predictions still suffer from large uncertainties, mainly due to the dependence on the factorisation and renormalisation scales.

A measurement of the $b$-hadron production cross section in proton-proton collisions at the Large Hadron Collider (LHC) provides a further test of QCD calculations for heavy-quark production at higher centre-of-mass energies. Recently the LHCb experiment measured the \bbbar{} and $B^+$~\cite{lhcb1,lhcb2,lhcb3} production cross sections in the forward region at $\sqrt{s}=7\,\TeV{}$, the CMS experiment measured the production cross sections for $B^+$, $B^0$, $B^0_s$ mesons, inclusive $b$-hadrons with muons, and $b\bar b$ decays with muons at $\sqrt{s}=7\,\TeV{}$~\cite{cms1,cms2,cms3,cms4,cms5}, and the ALICE experiment measured the \bbbar{} production cross section in $pp$ collisions at $\sqrt{s}=7\,\TeV{}$~\cite{alice}.

This paper presents a measurement of the $b$-hadron ($\Bhadrn$, a hadron containing a $b$-quark and not a $\bar b$-quark) production cross section at a centre-of-mass energy of 7 TeV with the ATLAS detector at the LHC, and its comparison with the NLO QCD theoretical predictions. 
The measurement requires the partial reconstruction of the $b$-hadron decay final state $\dsmu X$, with the $\dstarp$ reconstructed through the fully hadronic decay chain $\dstarp \rightarrow \pi^+ D^0(\rightarrow K^- \pi^+)$. This sample was collected by ATLAS between August and October 2010 using events selected by a single-muon trigger, and corresponds to a total integrated luminosity of 3.3\,\ipb.

%% file: detector.tex
\section{The ATLAS detector}
\label{sec:det}

The ATLAS detector \cite{atlas} covers almost the full solid angle around the collision point with layers of tracking detectors, calorimeters and muon chambers. For the measurement presented in this paper, the inner detector tracking devices, the muon spectrometer and the trigger system are of particular importance.

The inner detector (ID) has full coverage in $\phi$ and covers the pseudorapidity range $|\eta|<2.5$. It consists of a silicon pixel detector, a silicon microstrip tracker and a transition radiation tracker composed of drift tubes. These detectors are located at radial distances of 50.5--1066\,mm from the interaction point and are surrounded by a thin superconducting solenoid providing a 2\,T axial magnetic field. The ID barrel consists of three layers of pixels, four double-layers of single-sided silicon microstrips, and 73 layers of drift tubes, while each ID end-cap has three layers of pixels, nine double-layers of single-sided silicon microstrips, and 160 layers of drift tubes.

The muon spectrometer covers the pseudorapidity range $|\eta|<2.7$ and is located within the magnetic field produced by three large superconducting air-core toroid systems. The muon spectrometer is divided into a barrel region ($|\eta|<1.05$) and two end-cap regions ($1.05<|\eta|<2.7$), within which the average magnetic fields are 0.5\,T and 1\,T respectively. Precise measurements are made in the bending plane by monitored drift tube chambers, or, in the innermost layer for $2.0<|\eta|<2.7$, by cathode strip chambers. Resistive plate chambers in the barrel and thin gap chambers at $|\eta|<2.4$ in the end-caps are used as trigger chambers. The chambers are arranged in three layers, such that high \pT{} muons traverse at least three stations with a lever arm of several metres. 

A three-level trigger system is used to select interesting events. The first level is hardware-based, and uses a subset of the detector information to reduce the event rate to a design value of at most 75~kHz. This is followed by two software-based trigger levels, together known as the high level trigger, which finally reduce  the event rate to about 200~Hz.

%% file: outline.tex
\section{Outline of the measurement}
\label{sec:outline}

The first result presented in this paper is the $\dsmucross X$ production cross section, measured in a limited fiducial acceptance for the $\dsmu$ final state. 
 Given the integrated luminosity $\mathscr{L}$ of the data sample, and the branching ratio $\mathscr{B}$ of the $\dstarp$ cascade decay $\dstarp\rightarrow \pi^+ D^0 (\rightarrow K^- \pi^+)$, the $\dsmucross X$ cross section is defined as:
\begin{equation}
\sigma(pp\to\dsmucrossext) = \frac{f_b \,N({\dstarp\mu^- + \dstarm\mu^+})}{2 \epsilon \mathscr{B}\mathscr{L}} 
\label{eq:eq1}
\end{equation}
where $N({\dstarp\mu^- + \dstarm\mu^+})$ is the total number of reconstructed candidates, $f_b$ is the fraction of candidates originating from the decay $\dsmucross X$ and $\epsilon$ is the signal reconstruction efficiency. The efficiency takes into account reconstruction and muon trigger efficiencies, including the loss of events where the $\dstarp$ falls within the fiducial acceptance, but the decay products ($\pi$ or $K$) can not be reconstructed because they fall outside the $\pT{}$ and $\eta$ acceptance. The number $N$ of reconstructed candidates includes both  $\dstarp\mu^-$ and $\dstarm\mu^+$ combinations: assuming that $b$- and $\bar b$-quarks are produced with the same rate at the LHC, the factor of two is needed to quote the cross section for hadrons containing a $b$-quark.
The value of the branching ratio $\mathscr{B}$ can be obtained by combining the world average values of the branching ratios $\dstarp\rightarrow \pi^+ D^0$ and $D^0\rightarrow K^- \pi^+$~\cite{PDG}, and is $(2.63\pm0.04)\,\%$.

The parameters $N$, $f_b$ and $\epsilon$ are determined as functions of the transverse momentum and pseudorapidity of the $\dsmu$ pairs, in order to measure the differential cross sections. The detailed calculation of these parameters is discussed in the following sections.

To obtain the $b$-hadron production cross section $\sigma(pp\to\Bcross\,X)$, the $\dsmucross X$ cross section is divided by an acceptance correction $\alpha$, accounting for the fiducial region in which this is measured, and by the inclusive branching ratio $\BR$. For this branching ratio the world average value is $(2.75\pm0.19)\,\%$, assuming the world average values of the $b$-hadronisation fractions~\cite{PDG}. The dominant contributions to the sample are from $B^0$ mesons, through the decay $B^0\to \dstarm \mu^+ \nu_\mu$ and its charge conjugate.

%% file: montecarlo.tex
\section{Event simulation and NLO cross section predictions}
\label{sec:mc}

Monte Carlo (MC) simulated samples are used to optimise the selection criteria (Section~\ref{sec:rec}) and to evaluate the $\dsmu$ signal composition and reconstruction efficiency (Sections~\ref{sec:sample} and \ref{sec:eff}). The different $b$- and $c$-quark sources of $\dsmu$ are studied using inclusive samples of \bbbar{} and \ccbar{} events having at least one muon with $\pT>4\GeV$ and $|\eta|<2.5$ in the final state. Both samples are generated with {\sc Pythia}~\cite{pythia}, using the ATLAS AMBT1 tuning~\cite{atlastune}. The ATLAS detector response to the passage of the generated particles is simulated with {\sc Geant4}~\cite{geant, geant2}, and the simulated events are fully reconstructed with the same software used to process the collision data. 

To compare the measurements with theoretical predictions, NLO QCD calculations, matched with a leading-logarithmic parton shower MC simulation, are used. Predictions for \bbbar{} production at the LHC at $\sqrt{s}=7\TeV$ are evaluated with two packages: {\sc mc@nlo} 4.0~\cite{mcnlo1,mcnlo2} and {\sc Powheg-hvq} 1.01~\cite{pow1,pow2}. {\sc mc@nlo} is matched with the {\sc Herwig} 6.5~\cite{herwig1} MC event generator, while {\sc Powheg} is used with both {\sc Herwig} 6.5 and {\sc Pythia} 6.4~\cite{pythia}. For all the predictions, the inclusive branching ratio $\BR$ is set to the world average value.

The following set of input parameters is used to perform all theoretical predictions:
\begin{itemize}
	\item CTEQ6.6~\cite{cteq} parameterisation for the proton parton distribution function (PDF).
	\item $b$-quark mass $m_b$ of 4.75\,\GeV{}~\cite{PDG}. 
	\item Renormalisation and factorisation scales set to $\mu_r = \mu_f = \mu$, where $\mu$ has different definitions for {\sc mc@nlo} and {\sc Powheg}. For {\sc mc@nlo}:
	\begin{linenomath}$$ \mu^2 = m^2_Q + \frac{(p_{\mathrm{T},Q}+p_{\mathrm{T},\bar Q})^2}{4}$$\end{linenomath}
	where $p_{\mathrm{T},Q}$ and $p_{\mathrm{T},\bar Q}$ are the transverse momenta of the produced heavy quark and antiquark, and $m_Q$ is the heavy-quark mass. For {\sc Powheg}:
	\begin{linenomath}$$ \mu^2 = m^2_Q + (m^2_{Q\bar Q}/4 - m^2_Q) \sin^2(\theta_Q)$$\end{linenomath}
	where $m_{Q\bar Q}$ is the invariant mass of the $Q\bar Q$ system and $\theta_Q$ is the polar angle of the heavy quark in the $Q\bar Q$ rest frame.
	\item Heavy-quark hadronisation: cluster model~\cite{cluster} for {\sc Herwig}; Lund string model~\cite{lund} with Bowler modification~\cite{bowler} of the Lund symmetric fragmentation function~\cite{lund2} for {\sc Pythia}.
\end{itemize} 

The following sources of theoretical uncertainties are included in the NLO predictions:
\begin{itemize}
	\item Scale uncertainty, determined by varying $\mu_r$ and $\mu_f$ independently to $\mu/2$ and $2\mu$, with the additional constraint $1/2 < \mu_r/\mu_f < 2$, and selecting the largest positive and negative variations.
	\item $m_b$ uncertainty, determined by varying the $b$-quark mass by $\pm$0.25\,\GeV.
	\item PDF uncertainty, determined by using the CTEQ6.6 PDF error eigenvectors; the total uncertainty is obtained by varying each parameter independently within these errors and summing the resulting variations in quadrature.
	\item Hadronisation uncertainty, determined in {\sc Pythia} by using the Peterson fragmentation function~\cite{peterson} instead of the Bowler one, with extreme choices of the $b$-quark fragmentation parameter: $\epsilon_b=0.002$ and $\epsilon_b = 0.01$.
\end{itemize}

In addition to the final comparison with the experimental measurement, these theoretical predictions are used to unfold and extrapolate the measured cross sections (Sections~\ref{sec:unfold} and \ref{sec:acc}), and to extrapolate to the full kinematic phase space (Section~\ref{sec:discussion}).
In the following, {\sc Powheg+Pythia} is used as the default prediction.

%% file: selection.tex
\section{Data selection and reconstruction  of the $\dsmu$ decay}
\label{sec:rec}

The $\dsmu$ (including its charge conjugate) sample was collected during stable proton-proton collisions.
 Events were selected by a single-muon trigger, which requires a muon, reconstructed by the high level trigger, with $\pT>6\GeV$. This trigger was prescaled during the last part of the 2010 data-taking period. Taking into account the prescale factors, this data sample corresponds to an integrated luminosity of 3.3\,\ipb.
  
 The $\dstarp$ candidates are reconstructed through the fully hadronic decay chain $\dstarp\to \pi^+ D^0 (\to K^- \pi^+)$, using only good quality tracks, i.e. tracks with at least five silicon detector hits, and at least one of them in the pixel detector.
 
 The $b$-hadron and $D^0$ decay vertices are reconstructed and fitted simultaneously.  To perform the vertexing, an iterative procedure based on a fast Kalman filtering method is used. This allows to reconstruct consecutively all the vertices of the same decay chain, using the full information from track reconstruction (particles trajectories with complete error matrices).
    All pairs of opposite charge particle tracks are fitted to a single vertex to form $D^0$ candidates, assigning to each track, in turn,
the kaon or the pion mass, with the additional requirement $\pT>1\GeV$ for both the kaon and pion candidate; the resulting $D^0$ candidate is reconstructed by combining the kaon and pion four-momenta. The $D^0$ path is then extrapolated back and fitted with a track of opposite charge to the candidate kaon, requiring $\pT>250\MeV$ and assigning to it the pion mass, to form the $\dstarp$ candidate, and with a muon with $\pT>6\GeV$ and $|\eta|<2.4$ to form the $b$-hadron vertex. No requirements are made here on the muon charge; only opposite charge combinations $\dsmu$ are used in the analysis, while same charge combinations are used to cross-check the background. The muon is also required to have fired the trigger. To ensure good fit quality, the global $\chi^2$ probability of the combined fit must satisfy $P(\chi^2)>0.001$. To avoid an additional systematic uncertainty no requirement on the $b$-hadron vertex decay length is applied.

The $\dstarp$ candidate is accepted if it satisfies $\pT(K^-\pi^+\pi^+) >4.5\GeV$ and $|\eta(K^-\pi^+\pi^+)|<2.5$, and either (a) $|m({K^-\pi^+})-m({D^0})|<64\MeV$ in the region $\pT(K^-\pi^+\pi^+)>12\GeV$ and $|\eta(K^-\pi^+\pi^+)|>1.3$, or (b) $|m(K^-\pi^+)-m(D^0)|<40\MeV$ elsewhere. Here $m(D^0)$ is the world average value for the $D^0$ mass~\cite{PDG}.      This last selection cut is divided into two different kinematic regions due to the changing $D^0$ mass resolution. The $\dsmu$ candidate must have an invariant mass in the range 2.5--5.4\,\GeV. The upper invariant mass cut matches the physical requirement of not exceeding the mass of the $B$-mesons. 
 
\begin{figure}[!h]
  \subfigure[]{
    \includegraphics[angle=0,width=0.495\columnwidth]{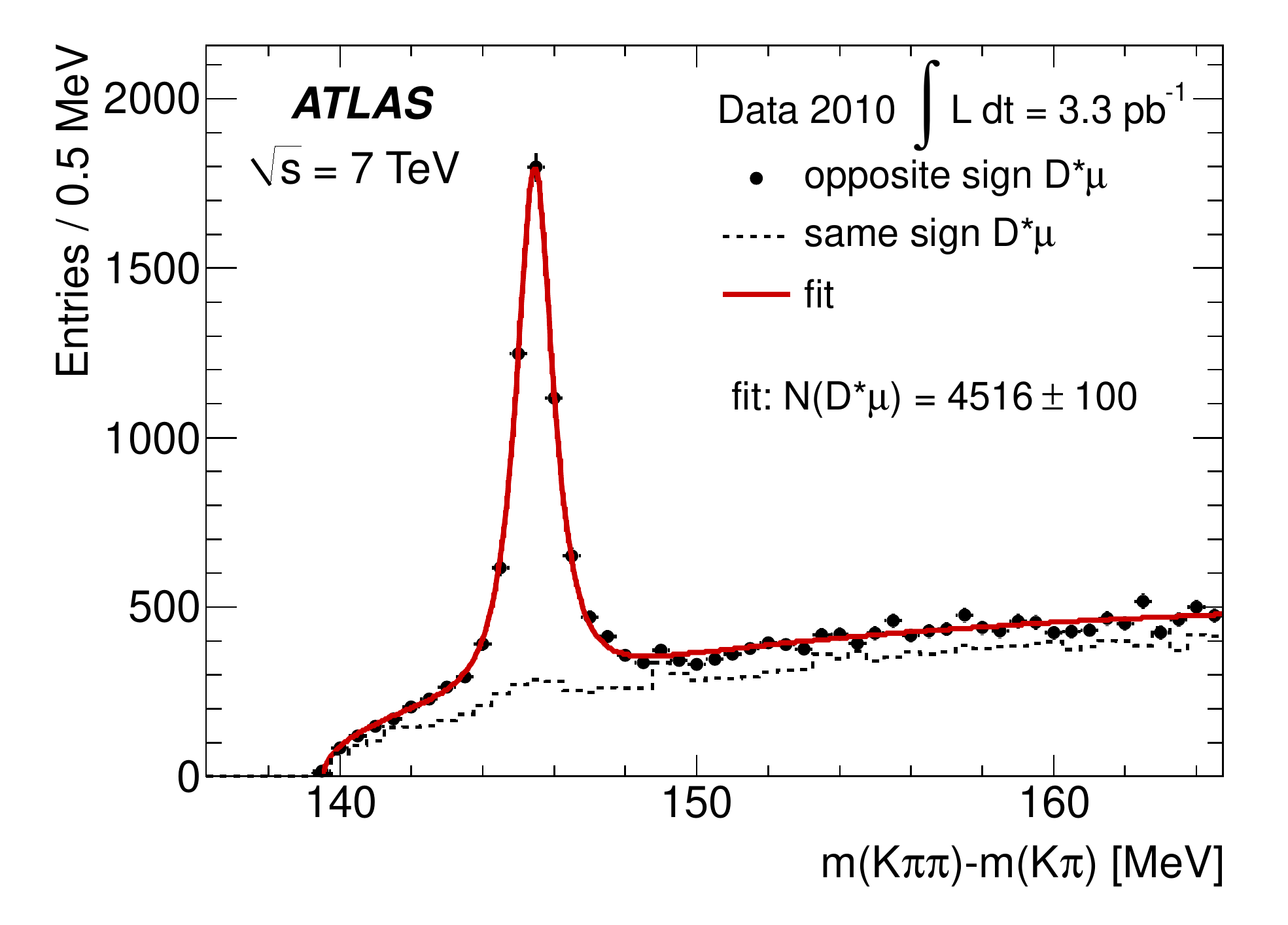}
  }
    \subfigure[]{
       \includegraphics[angle=0,width=0.495\columnwidth]{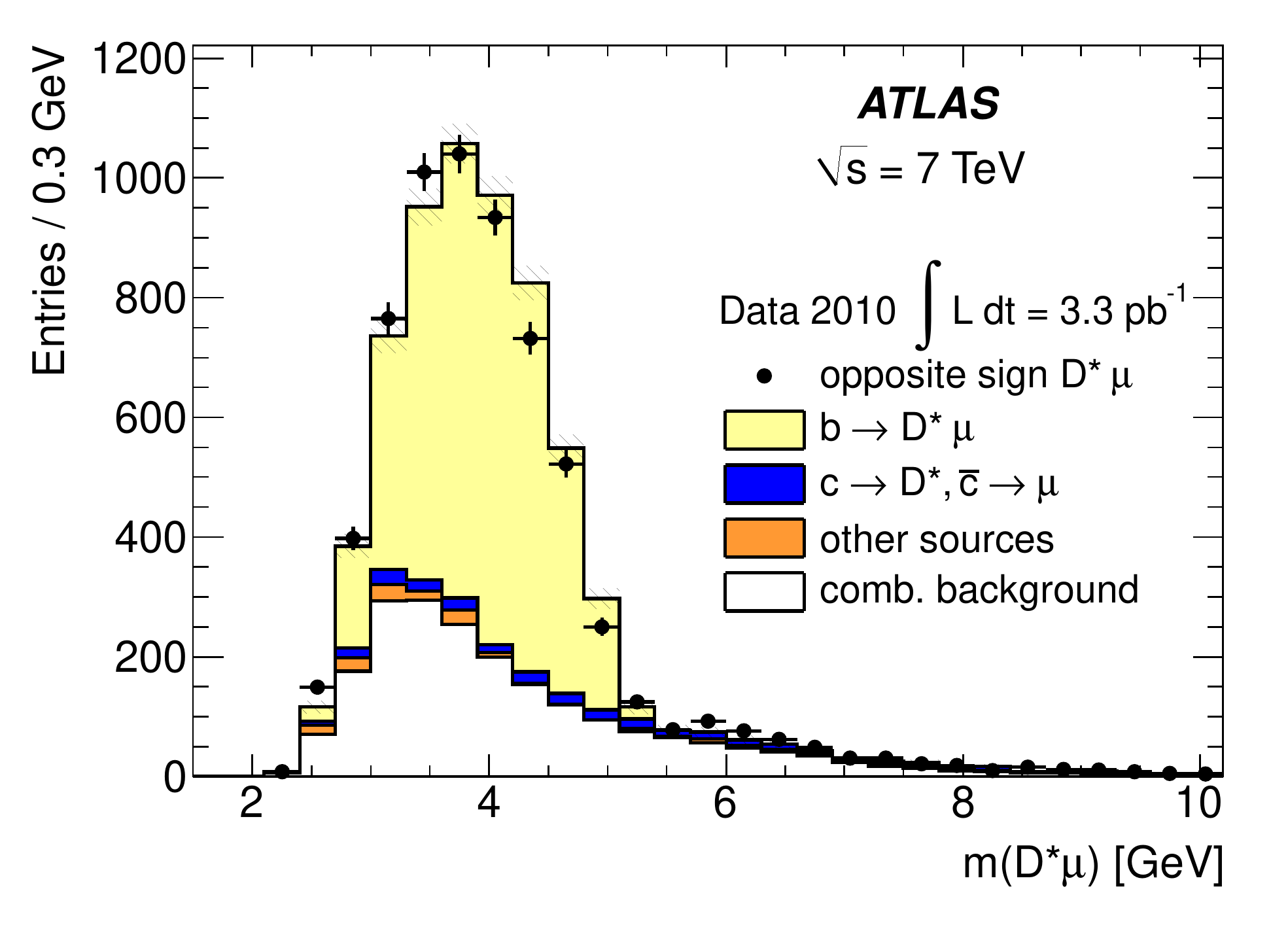}
 }
   \caption{(a) Distribution of the mass difference $\Delta m$ for $\dstaro\mu$ combinations of opposite charge (points) and same charge (dashed line). The solid line shows the result of the fit described in the text. (b)~Distribution~of the opposite charge $\dstaro\mu$ invariant mass, for mass combinations within $\pm3\sigma$ of the $\Delta m$ peak, without applying the invariant mass cut described in the text. The measured distribution is compared with the MC simulation, including the contribution of different sources of signal. The hashed bands show the MC statistical uncertainty.   }
\label{fig:dstar}
  \end{figure}

 Because of the kinematics of the $\dstarp$ decay, the prompt pion takes only a small fraction of the energy. The $\dstarp$
signal is therefore studied as a function of the mass difference $\Delta m$ between
the $\dstarp$ and $D^0$ candidates. Real $\dstarp$ mesons are expected to form a peak in $\Delta m$ around
145.4\,\MeV, while the combinatorial background gives a rising distribution,
starting at the pion mass. The combinatorial background is made of fake $\dsmu$ candidates, created from combinations of tracks which pass the selection cuts, but do not come from a $\dsmu$ signal.
Figure~\ref{fig:dstar} (a) shows a clear signal in the distribution of $\Delta m$ for the reconstructed opposite charge $D^*\mu$ pairs. The dashed histogram shows the corresponding $\Delta m$ distribution for the same charge combinations $\dstaro^{\pm}\mu^\pm$, showing a very small excess around 145.4\,\MeV, whose origin is described in Section~\ref{sec:sample}.

The opposite charged signal distribution is fitted using a modified Gaussian (${G}^\mathrm{mod}$), which provides a good description of the tails of the signal distribution. The modified Gaussian has the form:
\begin{equation}
 	{G}^\mathrm{mod}(x) \propto \mbox{exp}[-0.5\cdot x^{1+\frac{1}{1+0.5x}}]
\end{equation}
where $x=|(\Delta m - \Delta m_0)/\sigma|$ and $\Delta m_0$ and $\sigma$, free parameters in the fit, are the mean and width of the $\Delta m$ peak.

The combinatorial background is fitted with a power function multiplied by an exponential function: 
\begin{equation}
B(\Delta m) \propto (\Delta m-m_{\pi})^{\alpha}e^{-\beta(\Delta m-m_{\pi})} 
\end{equation}
where $\alpha$ and $\beta$ are free fit parameters, and $m_\pi$ is the charged pion mass.

The fitted yield is $4516\pm100$ events, with a fitted $\Delta m_0=145.463\pm0.015\MeV$, to be compared with the world average value $145.421\pm0.010\MeV$~\cite{PDG}, and a fitted $\sigma=0.49\pm0.03\MeV$. The uncertainties on the fitted $\Delta m_0$ and $\sigma$ values are statistical only.

 Figure \ref{fig:dstar} (b) shows the $\dsmu$ invariant mass distribution selected in a region of $3\sigma$ around the $\Delta m$ peak, without applying any $\dsmu$ invariant mass cut. The measured distribution is compared with the MC \bbbar{}+\ccbar{} simulation described in Section~\ref{sec:mc}, which takes into account the contribution of different physical sources to the $\dsmu$ signal, as discussed in more detail in Section~\ref{sec:sample}. The MC simulation is separately normalised to the number of signal and background events in data. The selection on $m(\dsmu)$ has full efficiency for the signal, while rejecting part of the combinatorial background and physical processes other than a single $b$-hadron decay.
  
 In order to evaluate differential cross sections, the sample is divided into six $\pT(\dsmu)$ bins and five $|\eta(\dsmu)|$ bins.
The $\Delta m$ distribution in each bin is fitted independently using the same fitting procedure as for the total sample. The number of candidates in each bin is reported in Table~\ref{tab:ptetabins}, together with its statistical uncertainty from the fit. 

\ifthenelse{\boolean{supp}}{The total numbers $N(\dsmu)$ obtained as sums of the binned yields differ slightly   
from the yield obtained from the total sample fit, because of the independent fitting procedures.}{}

\input{tab_01}

%% file: tab_01.tex
\begin{table}[h]
\centering
\subtable{
\begin{tabular}{|c|l|}
\hline
  $\pT(\dsmu)$  &  $N(\dsmu)$  \\
 \hline
 \hline
 $\phantom{0}$9--12 \GeV &   $\phantom{0}334 \pm 33$  \\
 \hline
 12--15 \GeV               &   $1211 \pm 56$  \\ 
 \hline
  15--20 \GeV &   $1527 \pm 55 $  \\
 \hline
 20--30 \GeV               &   $1049 \pm 42$  \\ 
 \hline
 30--45 \GeV &   $\phantom{0}310 \pm 21$  \\
 \hline
 45--80 \GeV               &   $\phantom{00}76 \pm 10$ \\ 
 \hline
\end{tabular}
}\qquad\qquad
\subtable{
\begin{tabular}{|c|l|}
\hline
  $|\eta(\dsmu)|$   &  $N(\dsmu)$  \\
 \hline
 \hline
 0.0--0.5              &   $1330 \pm 47$  \\
 \hline
 0.5--1.0               &   $1207 \pm 47 $  \\ 
 \hline
  1.0--1.5             &   $\phantom{0}919 \pm 48$  \\
 \hline
 1.5--2.0               &   $\phantom{0}890 \pm 60$  \\ 
 \hline
 2.0--2.5              &   $\phantom{0}317 \pm 37$  \\
 \hline
\end{tabular}
}
\caption{Fitted number of opposite charge $D^*\mu$ pairs for different $\pT$ and $|\eta|$ bins.}
\label{tab:ptetabins}
\end{table}

%% file: samplecomp.tex
\section{ $\dsmu$ sample composition}
\label{sec:sample}

Various processes contribute to the $\dsmu$ data sample:
\begin{itemize}
	\item Direct semileptonic decay: $b\rightarrow \dstarp\mu^- X$; this is the {signal} contribution used for this measurement.
	 \item Decays of two $c$-hadrons, one of them decaying semileptonically: $c\rightarrow \dstarp X, \,\bar c \rightarrow \mu^- X'$.
	\item Direct semileptonic $\tau$ decay: $b\rightarrow \dstarp\tau^- X, \,\tau^-\rightarrow \mu^- \bar\nu_\mu \nu_\tau(\gamma)$.
	\item Decays of $b$-hadrons with two $c$-hadrons in the final state, one of them decaying semileptonically: $b\rightarrow \dstarp \bar D X,\, \bar D \rightarrow \mu^- X'$.

	\item Decays of two $b$-hadrons, one of them decaying semileptonically: $b\rightarrow \dstarp X, \,\bar b \rightarrow \mu X'$. This source contributes to opposite-sign and same-sign charge combinations,
    depending on the direct or indirect semileptonic decay relative branching
    ratio and on the neutral $b$-meson oscillation rate. This explains the small excess observed in Figure~\ref{fig:dstar} (a) in the peak region of the same sign charge $\Delta m$ distribution.

	\item A $\dstarp$ meson accompanied by a fake muon, contributing to both opposite-sign and same-sign charge combinations. The contribution from combinations with misidentified muon charge is negligible.
\end{itemize}

For the purposes of this measurement, only the direct semileptonic component is of interest. Therefore it is necessary to evaluate the fraction of the reconstructed $\dsmu$ sample that actually originates from direct semileptonic $b$ decays. This is estimated from the MC simulation. 
The most significant $\dsmu$ contributions are listed in Table \ref{tab:comptot}, together with the MC statistical uncertainty.

\input{tab_02}

The fractions from single $b$ semileptonic decays $f_b$, 
evaluated in the various $\pT$ and $|\eta|$ bins of the $\dsmu$ pair, are reported in Table~\ref{tab:comp}, together with the MC statistical uncertainty of the calculations. These values are used for the differential cross section measurements.

\input{tab_03}

%% file: tab_02.tex
\begin{table}[!h]
\renewcommand{\arraystretch}{1.1}
\centering
\subtable{
\begin{tabular}{|l|c|}
\hline
  Source  &  Fraction (\%)  \\
 \hline
 \hline
 $b\rightarrow \dstarp\mu^-X$ &   $93.2 \pm 0.3 $  \\
 \hline
 $c\rightarrow \dstarp X, \,\bar c\rightarrow \mu^- X'$           &   $\phantom{0} 3.8\pm0.2$  \\ 
 \hline
  $b\rightarrow \dstarp \tau^- X,\, \tau^-\rightarrow\mu^- X'$ &   $\phantom{0} 1.5\pm0.1$  \\
 \hline
 $b\rightarrow \dstarp\bar D X,\,\bar D\rightarrow\mu^- X'$              &   $\phantom{0}0.9\pm0.1$  \\ 
 \hline
 others              &   $\phantom{0}0.6\pm0.1$  \\ 
 \hline
 \end{tabular}
}
\caption{Different sources contributing to the $\dsmu$ sample. The uncertainties are due to MC statistics.}
\label{tab:comptot}
\end{table}

%% file: tab_03.tex
\begin{table}[!h]
\centering
\subtable{
\begin{tabular}{|c|c|}
\hline
  $\pT(\dsmu)$  &  $f_b\,(\%)$  \\
 \hline
 \hline
 $\phantom{0}$9--12 \GeV &   $90.8 \pm 1.2 $  \\
 \hline
 12--15 \GeV               &   $92.7 \pm 0.5  $  \\ 
 \hline
  15--20 \GeV &   $93.8 \pm 0.4  $  \\
 \hline
 20--30 \GeV               &   $93.2 \pm 0.5 $  \\ 
 \hline
 30--45 \GeV &   $93.8 \pm 0.9 $  \\
 \hline
 45--80 \GeV               &   $93.1 \pm 1.9  $ \\ 
 \hline
\end{tabular}
}\qquad\qquad
\subtable{
\begin{tabular}{|c|c|}
\hline
 $|\eta(\dsmu)|$   &  $f_b\,(\%)$  \\
 \hline
 \hline
 0.0--0.5 &   $93.0 \pm 0.5 $  \\
 \hline
 0.5--1.0               &   $92.6 \pm 0.5  $  \\ 
 \hline
  1.0--1.5&   $93.4 \pm 0.6  $  \\
 \hline
 1.5--2.0               &   $93.5 \pm 0.6 $  \\ 
 \hline
 2.0--2.5 &   $94.6 \pm 0.9 $  \\
 \hline
\end{tabular}
}
\caption{Fractions of single $b$ semileptonic decays in different $\pT(\dsmu)$ and $|\eta(\dsmu)|$  bins. The uncertainties are due to MC statistics.}
\label{tab:comp}
\end{table}

%% file: efficiency.tex
\section{Reconstruction and muon trigger efficiency}
\label{sec:eff}

The overall efficiency $\epsilon$ for $\dsmucross X$ decays to enter the $\dsmu$ sample, which includes the reconstruction, muon trigger and selection efficiencies, is evaluated as a product of three different components, in order to combine MC and data-driven efficiency calculations. Since the only requirement is the single $b$ detection efficiency, the \bbbar{} MC sample is used. The components are defined as:
\begin{equation}
\epsilon_\mathrm{reco}  = \frac{N(\mbox{true } \dsmu \mbox{ with $\mu$ and tracks reconstructed} )}{N(\mbox{true } \dsmu)}
\end{equation}
\begin{equation}
\epsilon_\mathrm{trigger}  = \frac{N(\mbox{true } \dsmu \mbox{ with $\mu$ and tracks reconstructed, $\mu$ matched to trigger)}}{N(\mbox{true }\dsmu \mbox{ with $\mu$ and tracks reconstructed} )}
\end{equation}
\begin{equation}
\epsilon_\mathrm{selection}  = \frac{N(\mbox{true }\dsmu \mbox{ with $\mu$ and tracks rec., $\mu$ matched to trigger, $\dsmu$ selection})}{N(\mbox{true }\dsmu \mbox{ with $\mu$ and tracks rec., $\mu$ matched to trigger})}
\end{equation}
where the number of true $\dsmu$ pairs is calculated within the fiducial kinematic region $\pT(\dstarp)>4.5\GeV$, $\pT(\mu^-)>6\GeV$, $|\eta(\dstarp)|<2.5$ and $|\eta(\mu^-)|<2.4$.
Events where the $\dstarp$ is inside the fiducial region, but its decay products are not fully reconstructed, contribute to $\epsilon_\mathrm{reco}$.

Both $\epsilon_\mathrm{reco}$ and $\epsilon_\mathrm{selection}$ are taken from MC simulation. However $\epsilon_\mathrm{trigger}$, which is the fraction of the reconstructed muons that actually satisfied the trigger, is measured directly from data using $\Jpsi\rightarrow \mu^+\mu^-$ samples~\cite{jpsi}. These efficiencies are evaluated for the same data-taking periods used in this measurement. 

The overall efficiency $\epsilon$ is given by:
\begin{equation}
	\epsilon = \epsilon_\mathrm{reco} (\mbox{MC}) \epsilon_\mathrm{trigger}  (\mbox{data})  \epsilon_\mathrm{selection}  (\mbox{MC}) 
\end{equation} 

The different efficiency components, together with the related statistical uncertainties, are determined as $\epsilon_\mathrm{reco}=(48.3\pm0.4)\%$, $\epsilon_\mathrm{trigger}=(81.9\pm0.4)\%$ and $\epsilon_\mathrm{selection}=(79.1\pm0.5)\%$.
The overall efficiency is $(31.3\pm0.4)\%$, and the values obtained in $\pT(\dsmu)$ and $|\eta(\dsmu)|$ bins are reported in Table~\ref{tab:receff}. A complete description of the systematic uncertainties follows in Section~\ref{sec:syst}. 

\input{tab_04}

%% file: tab_04.tex
\begin{table}[h]
\centering
\subtable{
\begin{tabular}{|c|c|}
\hline
  $\pT(\dsmu)$  &  $\epsilon\,(\%)$  \\
 \hline
 \hline
 $\phantom{0}$9--12 \GeV &   $21.2 \pm 0.9$  \\
 \hline
 12--15 \GeV               &   $26.7 \pm 0.6 $  \\ 
 \hline
  15--20 \GeV              &   $32.1 \pm 0.6 $  \\
 \hline
 20--30 \GeV               &   $38.8 \pm 0.9$  \\ 
 \hline
 30--45 \GeV               &   $45.2 \pm 1.7$  \\
 \hline
 45--80 \GeV               &   $\phantom{}52 \pm 4 $ \\ 
 \hline
\end{tabular}
}\qquad\qquad
\subtable{
\begin{tabular}{|c|c|}
\hline
  $|\eta(\dsmu)|$   &  $\epsilon\,(\%)$  \\
 \hline
 \hline
 0.0--0.5               &   $37.5 \pm 0.7$  \\
 \hline
 0.5--1.0               &   $37.2 \pm 0.8 $  \\ 
 \hline
  1.0--1.5              &   $29.9 \pm 0.8 $  \\
 \hline
 1.5--2.0               &   $26.1 \pm 0.8$  \\ 
 \hline
 2.0--2.5               &   $16.1 \pm 0.9$  \\
 \hline
\end{tabular}
}
\caption{Overall efficiency $\epsilon$ for different $\pT(\dsmu)$ and $|\eta(\dsmu)|$ bins.}
\label{tab:receff}
\end{table}

%% file: systematics.tex
\section{Systematic uncertainties}
\label{sec:syst}

The uncertainty in the cross section due to each systematic variation is evaluated by repeating the entire analysis procedure and finding the change in the cross section value. The same strategy is adopted to evaluate bin-by-bin systematic uncertainties for the differential cross section measurements. 
The following sources are considered:

 \begin{itemize}
	\item Uncertainty of the yields from the fits, obtained by varying the fitting procedure in the following ways:
	\begin{itemize}
		\item reducing the high end of the $\Delta m$ range used for the $\dsmu$ signal fit by 4\,\MeV{}, from 165\,\MeV{} to 161\,\MeV{};
		\item changing the background parameterisation function to be $\propto 1- \mbox{exp}(-\alpha(\Delta m - m_{\pi})^{\beta})$, where $\alpha$ and $\beta$ are free fit parameters, which provides a $P(\chi^2)$ for the fit similar to that with the default background parameterisation.
	\end{itemize}
	\item Uncertainty of the sample composition estimate: the $f_b$ measurement depends on the $b/c$ cross section ratio used in the MC sample. 
	The ratio of the beauty and charm contributions to the inclusive $D^{*+}$ production, estimated using the life-time information, has been found to be in agreement with the ratio in {\sc Pythia}, within experimental uncertainties. To cover the uncertainties, the MC $b/c$ ratio is varied between 50\% and 200\% of its nominal value.
	\item Uncertainties of the muon trigger efficiencies are estimated from $J/\psi\rightarrow\mu^+\mu^-$ studies \cite{jpsi}.
	\item Uncertainty on tracking and muon reconstruction efficiency: the uncertainty on ID tracking efficiency is dominated by the detector material description used in MC simulations. This uncertainty is evaluated in studies of minimum bias events \cite{atlastune}. The muon reconstruction uncertainty is evaluated on $Z\to \mu^+\mu^-$ data samples~\cite{jpsi}. This systematic uncertainty is dominated by the ID tracking uncertainty.
	\item Model dependence of the reconstruction efficiency: the efficiency calculation could be affected by differences between the $\pT(\dsmu)$ and $\eta(\dsmu)$ spectra in data and MC simulation. To estimate the systematic uncertainty, the MC distribution is varied, while preserving consistency with the observed data distribution, and the resulting change in efficiency is computed after each variation.
	\item Uncertainty due to differences in the fit of the $D^0$ and $b$-hadron vertices between data and MC simulation: to estimate the systematic uncertainty, the MC $P(\chi^2)$ distribution is varied, while preserving consistency with the observed data distribution, and the resulting change in efficiency is computed after each variation.
	\item Uncertainty of the difference in $D^0$ mass resolution between data and MC simulation:  the efficiency calculation is corrected to account the difference between $D^0$ mass resolution in data and MC simulation. To estimate the systematic uncertainty, the error on the data-to-MC ratio of $D^0$ mass widths is propagated to the efficiency. 
	\item NLO prediction uncertainty: since the NLO predictions are also used as an active part of the analysis for unfolding (Section~\ref{sec:unfold}) and acceptance corrections (Section~\ref{sec:acc}), the theoretical uncertainties and the use of different predictions introduce additional systematic uncertainties to the experimental measurements. These are evaluated by repeating the entire analysis, introducing different theoretical uncertainties (Section~\ref{sec:mc}) to the default central prediction ({\sc Powheg+Pythia}), and using a different theoretical prediction ({\sc Powheg+Herwig} and {\sc mc@nlo}): positive and negative differences obtained with respect to using the central prediction are separately summed in quadrature. The use of the predictions matched with {\sc Herwig} produces visible asymmetries in the uncertainties of the acceptance corrections (Section~\ref{sec:acc}).
	\item Uncertainty of the luminosity measurement ($\pm 3.4\%$)\cite{lumi1,lumi2}. 
	\item Relative uncertainty on the branching fractions of the different decay chains, obtained from the world averages \cite{PDG}: $b\rightarrow \dstarp\mu^- X \,(\pm 7\,\%), \dstarp\rightarrow D^0 \pi^+ \,(\pm 0.7\,\%), \, D^0\rightarrow K^-\pi^+ \,(\pm 1.3\,\%)  $.

\end{itemize}

In Sections~\ref{sec:unfold} and \ref{sec:acc}, tables are shown with these uncertainties quoted after each step of the analysis. 

%% file: unfolding.tex
\section{Differential cross sections for $\dsmucross X$ production}
\label{sec:unfold}

Differential cross sections for $\dsmucross X$ production as a function of the $\pT$ and $|\eta|$ of the $\dsmu$ pairs are evaluated by using Equation~\ref{eq:eq1} and dividing by the bin width. The results are shown in Table~\ref{tab:Xdsmu}.  

\input{tab_05}

To extract differential cross sections as a function of the $\pT$ and $|\eta|$ of the $b$-hadron, it is necessary to correct the observed $\pT(\dsmu)$ and $|\eta(\dsmu)|$ distributions using Monte Carlo simulations, in order to take into account the kinematics of the missing particles from the decay $\dsmucross X$. This procedure is known as unfolding~\cite{unfold1,unfold2,unfold3,unfold4}.
The unfolding approach used in this paper is based on the iterative method described in Ref.~\cite{bayesunf}, containing elements of Bayesian statistics. 

The element $F_{ij}$ of the response matrix $F$ for a $b$-hadron in a $\pT/|\eta|(\Bhadrn)$ bin $j$ to decay into a $\dsmu$ in $\pT/|\eta|(\dsmu)$ bin $i$ can be interpreted as a conditional probability
\begin{equation}
 F_{ij} = P(\dsmu \mbox{ in bin }i |   \Bhadrn \mbox{ in bin }j).  
\end{equation}
Given an initial set of probabilities $p_i$ for $b$-hadrons to be found in bin $i$, using Bayes' theorem one can obtain the expected number of $b$-hadrons in bin $i$, given a measured $\dsmu$ distribution:
\begin{equation}
N^{\Bhadrn}_i = \sum_{j=1}^{N_\mathrm{bin}} P(  \Bhadrn \mbox{ in bin }i | \dstaro\mu \mbox{ in bin }j) N^{\dsmu}_j = 
\sum_{j=1}^{N_\mathrm{bin}} \bigg( \frac{F_{ji}p_i}{\sum_k F_{jk}p_k}\bigg) N^{\dsmu}_j .
\end{equation}
A NLO Monte Carlo sample generated with {\sc Powheg+Pythia} is used to create the default response matrix $F$ and the initial prior probabilities $p$. The procedure is repeated with different MC generators, in order to evaluate systematic uncertainties.

The procedure can be iterated, taking as new prior probabilities the solutions of the previous step, i.e. $p_i=N^{\Bhadrn}_i / N^{\Bhadrn}_\mathrm{tot}$. 
After a large number of iterations, the procedure converges on the results obtained with a direct inversion of the response matrix $F$
\begin{equation}
N^{\Bhadrn}_i = \sum_{j=1}^{N_\mathrm{bin}} (F^{-1})_{ij} N^{\dsmu}_j.
\end{equation}
This method is known to be sensitive to statistical fluctuations~\cite{unfold1}, but this effect can be mitigated in the Bayesian method by truncating the procedure after a few iterations. 

The number of iterations was therefore optimised in Monte Carlo simulations with test measurements, comparing the values obtained after each iteration to the values expected from the MC-generated information, using a $\chi^2$ test. Two iterations are the optimal solution in this case, providing compatible results even when the response matrix $F$ and the prior probabilities $p$ are generated using different theoretical distributions.

The inversion method and the Bayesian method with a different number of iterations were employed as a check. Within the systematic uncertainties, all the results were found to be in agreement with the chosen default procedure.

A bias could occur in this procedure due to the possible mismodelling of the $\Bhadrn$ decays (e.g. $D^{**}$ decays contributing to the missing particles in the final state) in the simulation. It was verified with the simulation that the relevant $\dsmu$ kinematic variables have a small dependence on the specific $b$-hadron decay, and that a mismodelling of the $D^{**}$ branching ratios does not produce a significant effect. This is expected since the dominant $\dsmu$ contribution arises from direct $B^0$ decays without an intermediate $D^{**}$. 

Once the $\Bhadrn$ distribution is obtained, the differential $\dsmucross X$ cross sections are determined as a function of $\pT$ and $|\eta|$ of the $b$-hadron, inside the kinematic region $\pT(\dstarp)>4.5\GeV$, $\pT(\mu^-)>6\GeV$, $|\eta(\dstarp)|<2.5$ and $|\eta(\mu^-)|<2.4$.

\begin{figure}[h]
  \subfigure[]{
  \includegraphics[angle=0,width=0.495\columnwidth]{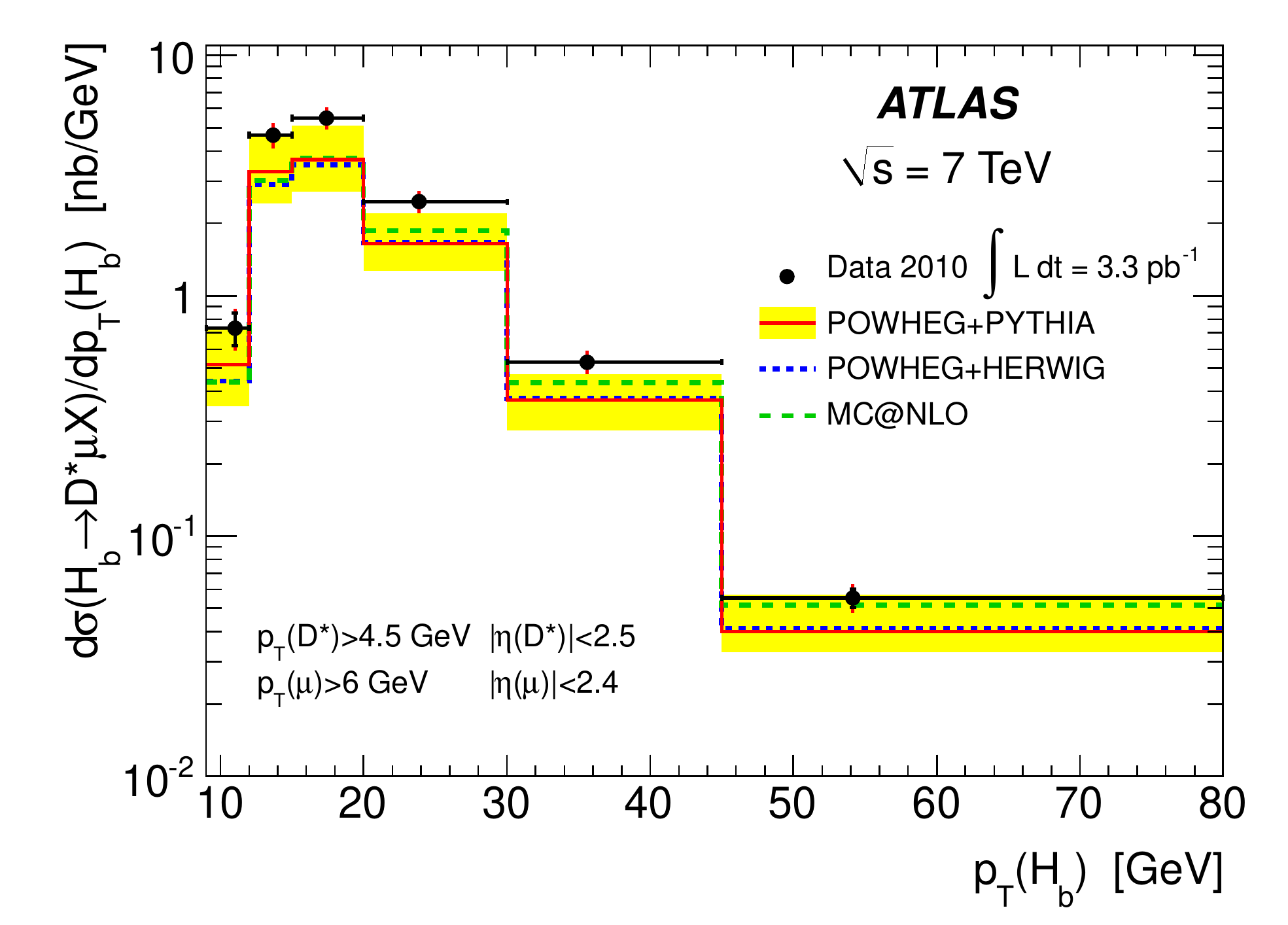} 
     }
    \subfigure[]{
    \includegraphics[angle=0,width=0.495\columnwidth]{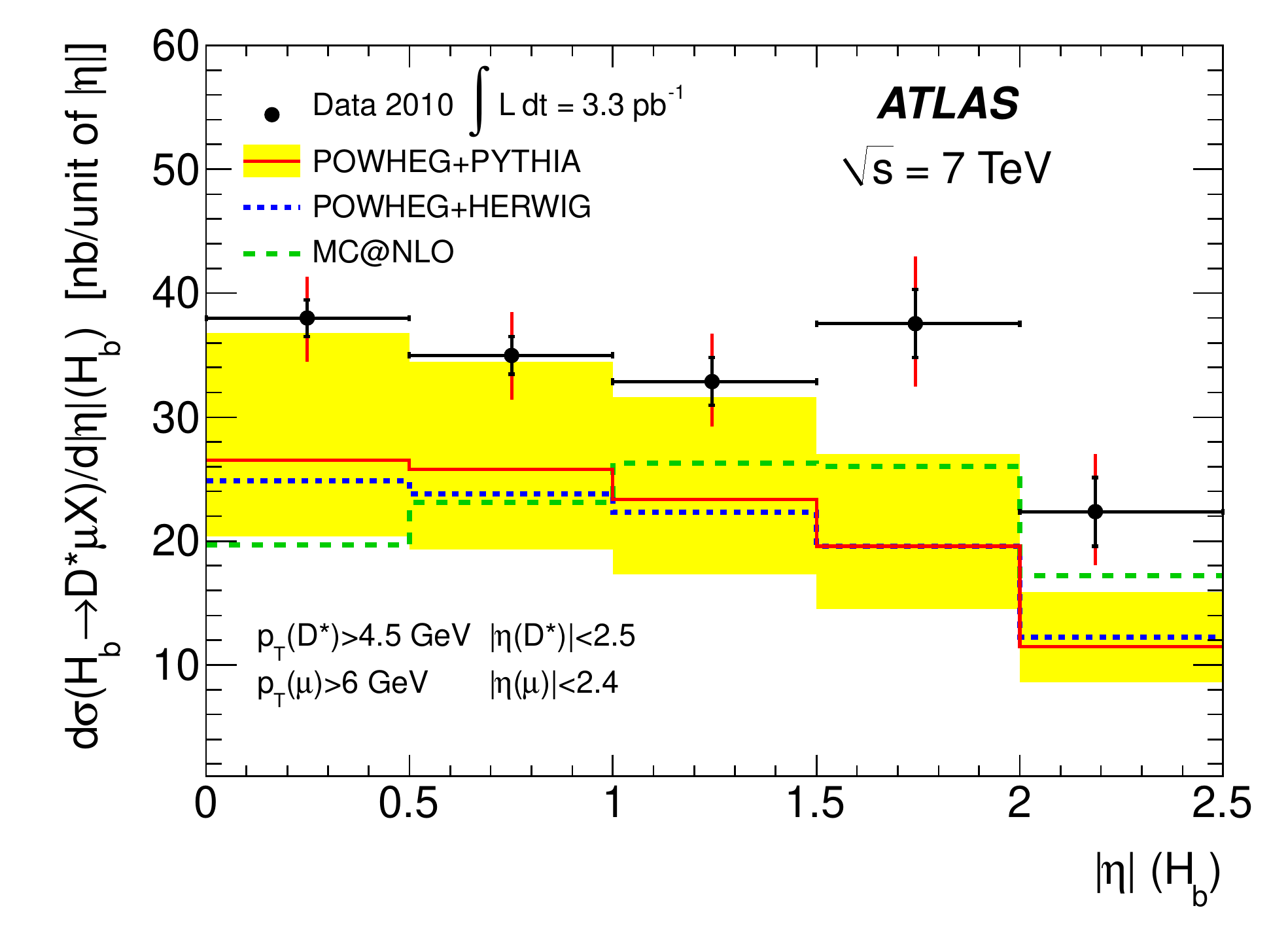}
  }
    \caption{Differential cross section for $\dsmucross X$ production as a function of (a) $\pT$ and (b) $|\eta|$ of the $b$-hadron, in the fiducial kinematical region $\pT(\dstarp)>4.5\GeV, \pT(\mu^-)>6\GeV, |\eta(\dstarp)|<2.5$ and $|\eta(\mu^-)|<2.4$. The measurement is compared with the theoretical predictions, as described in the text. The inner error bars of the data points are statistical uncertainties, the outer are statistical+total systematic uncertainties. 
 }
\label{crossbdsmu}
  \end{figure}

Figure~\ref{crossbdsmu} shows the measured differential cross sections, with comparisons to the NLO theoretical predictions. The {\sc Powheg+Pythia} shaded band refers to the total theoretical uncertainty of the prediction. The differential cross section values are reported in Table \ref{tab:Xsections}, together with the statistical and total systematic uncertainties. The individual contributions to the systematic uncertainties are listed in Tables \ref{syst1} and \ref{syst2}.
The comparison with data shows that NLO calculations underestimate the cross section, although the difference is within the combined experimental and theoretical uncertainties.

\input{tab_06}
\input{tab_07}
\input{tab_08}

The integrated $\dsmucross X$ cross section, inside the kinematic region $\pT(\dstarp)>4.5\GeV$, $\pT(\mu^-)>6\GeV$, $|\eta(\dstarp)|<2.5$ and $|\eta(\mu^-)|<2.4$, is:
\begin{linenomath}
$$\sigma(pp\to\dsmucrossext ) = 78.7 \pm 2.0\stat \pm 7.3\syst \pm 1.2(\mathscr{B}) \pm 2.7(\mathscr{L})  \,\,\mbox{nb}$$
\end{linenomath}

The integrated {\sc Powheg+Pythia} prediction, with its theoretical uncertainty, is:
\begin{linenomath}
$$\sigma(pp\to\dsmucrossext ) = 53 \, ^{+18}_{-12} (\mbox{scale}) \, ^{+3}_{-3} (m_b) \, ^{+3}_{-3} (\mbox{PDF}) \, ^{+6}_{-5} (\mbox{hadr.}) \,\,\mbox{nb}$$
\end{linenomath}

The corresponding {\sc Powheg+Herwig} prediction is 51\,nb, while {\sc mc@nlo} predicts 56\,nb, with similar theoretical uncertainties to the {\sc Powheg+Pythia} prediction.

%% file: tab_05.tex
\begin{table}[h]
\renewcommand{\arraystretch}{1.2}
\small
\centering
\subtable{
\begin{tabular}{|c|l|}
\hline
    $\pT(\dsmu)$  &  \multicolumn{1}{c|}{$\frac{\mathrm{d}\sigma(\dsmucross X)}{\mathrm{d}\pT(\dsmu)}$}  \\
  $[\mbox{\GeV}]$ & \multicolumn{1}{c|}{$[\mbox{nb/\GeV}]$} \\
 \hline
 \hline
 $\phantom{0}$9--12 &   $\phantom{00}2.78 \pm 0.29 {}^{+0.30}_{-0.30}$ \\ [0.0cm]
 \hline 
 12--15             &   $\phantom{000}8.2 \pm 0.4 {}^{+0.8}_{-0.8}$ \\ [0.0cm]
 \hline
  15--20 &             $\phantom{000}5.2 \pm 0.2 {}^{+0.5}_{-0.5} $  \\[0.0cm]
 \hline
 20--30              &   $\phantom{00}1.47 \pm 0.06 {}^{+0.15}_{-0.14}$ \\ [0.0cm]
 \hline
 30--45 &              $\phantom{0}0.250 \pm 0.018 {}^{+0.025}_{-0.024}$  \\[0.0cm]
 \hline
 45--80               &   $0.0229 \pm 0.0030 {}^{+0.0023}_{-0.0023}$ \\ [0.0cm]
 \hline
\end{tabular}
}\qquad\qquad
\subtable{
\begin{tabular}{|c|c|}
\hline
    $|\eta(\dsmu)|$   & $\frac{\mathrm{d}\sigma(\dsmucross X)}{\mathrm{d}|\eta(\dsmu)|}$ \\

                                  & $[\mbox{nb/unit of $|\eta|$}]$\\
 \hline
 \hline
 0.0--0.5 &  $38.4 \pm 1.5 {}^{+3.4}_{-3.4}$ \\
  \hline
 0.5--1.0  &         $34.9 \pm 1.4 {}^{+3.1}_{-3.1}$ \\
 \hline
  1.0--1.5&  $33.5 \pm 1.8 {}^{+3.4}_{-3.1}$ \\
 \hline
 1.5--2.0          &   $37.2 \pm 2.6 {}^{+4.7}_{-4.2}$ \\
 \hline
 2.0--2.5 & $21.7 \pm 2.6 {}^{+3.7}_{-3.1}$ \\
 \hline
\end{tabular}

}
  \caption{Differential cross sections for $\dsmucross X$ production as a function of $\pT$ and $|\eta|$ of the $\dsmu$ pair, in the fiducial kinematical region $\pT(\dstarp)>4.5\GeV, \pT(\mu^-)>6\GeV, |\eta(\dstarp)|<2.5$ and $|\eta(\mu^-)|<2.4$. The statistical and total systematic uncertainties are shown for each cross section.}
\label{tab:Xdsmu}
\end{table}

%% file: tab_06.tex
\begin{table}[h]
\renewcommand{\arraystretch}{1.2}
\centering
\subtable{
\begin{tabular}{|c|l|l|}
\hline
 $\pT(\Bhadrn)$  &  \multicolumn{1}{c|}{$\frac{\mathrm{d}\sigma(\dsmucross X)}{\mathrm{d}\pT(\Bhadrn)}$} & \multicolumn{1}{c|}{$\frac{\mathrm{d}\sigma(\Bcross X)}{\mathrm{d}\pT(\Bhadrn)}$} \\
  $[\mbox{\GeV}]$ & \multicolumn{1}{c|}{$[\mbox{nb/\GeV}]$} & \multicolumn{1}{c|}{$[\mbox{nb/\GeV}]$} \\
 \hline
 \hline
 $\phantom{0}$9--12 &   $\phantom{0}0.73 \pm 0.12 {}^{+0.09}_{-0.11}$ & $\phantom{0}(5.8 \pm 0.9 {}^{+0.8}_{-1.0})\cdot 10^3$\\
 \hline
 12--15             &   $\phantom{0}4.65 \pm 0.27 {}^{+0.50}_{-0.50}$  & $(2.37 \pm 0.14 {}^{+0.30}_{-0.33})\cdot 10^3$\\ 
 \hline
  15--20 &             $\phantom{0}5.48 \pm 0.19 {}^{+0.57}_{-0.54} $  & $\phantom{0}(9.1 \pm 0.3 {}^{+1.1}_{-1.1})\cdot10^2$\\
 \hline
 20--30              &   $\phantom{0}2.46 \pm 0.08 {}^{+0.26}_{-0.24}$ & $\phantom{A}212 \pm 7\, {}^{+26}_{-26}$\\ 
 \hline
 30--45 &              $0.530 \pm 0.025 {}^{+0.056}_{-0.062}$  & $\phantom{0}31.3 \pm 1.5 {}^{+3.9}_{-3.9}$\\
 \hline
 45--80               &   $0.055 \pm 0.005 {}^{+0.007}_{-0.006}$ & $\phantom{0}2.78 \pm 0.25 {}^{+0.38}_{-0.33}$\\ 
 \hline
\end{tabular}
}\qquad\qquad
\subtable{
\begin{tabular}{|c|c|c|}
\hline
$|\eta(\Bhadrn)|$   & $\frac{\mathrm{d}\sigma(\dsmucross X)}{\mathrm{d}|\eta(\Bhadrn)|}$ & $\frac{\mathrm{d}\sigma(\Bcross X)}{\mathrm{d}|\eta(\Bhadrn)|}$ \\
                        & $[\mbox{nb/unit of $|\eta|$}]$ & $[\mbox{$\mu$b/unit of $|\eta|$}]$ \\
 \hline
 \hline
 0.0--0.5 &                $38.0 \pm 1.5 {}^{+3.3}_{-3.3}$ & $14.3 \pm 0.6 {}^{+1.7}_{-2.7}$\\
  \hline
 0.5--1.0  &                $35.0 \pm 1.5 {}^{+3.2}_{-3.2}$ & $13.4 \pm 0.6 {}^{+1.8}_{-2.7}$\\
 \hline
  1.0--1.5&                 $32.9 \pm 1.9 {}^{+3.3}_{-3.1}$ & $13.1 \pm 0.7 {}^{+2.1}_{-2.9}$\\
 \hline
 1.5--2.0               &   $37.5 \pm 2.7 {}^{+4.7}_{-4.3}$ & $15.8 \pm 1.1 {}^{+2.4}_{-4.4}$\\
 \hline
 2.0--2.5 &                $22.3 \pm 2.8 {}^{+3.8}_{-3.2}$ & $13.3 \pm 1.6 {}^{+2.5}_{-4.5}$\\
 \hline
\end{tabular}
}
  \caption{Differential cross sections for $\dsmucross X$ and $\Bcross X$ production as a function of $\pT$ and $|\eta|$ of the $b$-hadron, in the fiducial kinematical regions $\pT(\dstarp)>4.5\GeV, \pT(\mu^-)>6\GeV, |\eta(\dstarp)|<2.5$ and $|\eta(\mu^-)|<2.4$, and $\pT(\Bcross)>9\GeV, |\eta(\Bcross)|<2.5$ respectively. The statistical and total systematic uncertainties are shown for each cross section.}
\label{tab:Xsections}
\end{table}

%% file: tab_07.tex
\begin{table}[!h]
\renewcommand{\arraystretch}{1.1}

\begin{center}
\begin{tabular}{|l||c|c|c|c|c|c|}
\hline
$\pT$ bin (\GeV)                          &       9--12      & 12--15     & 15--20      &  20--30    &  30--45       &  45--80 \\
 \hline
 \hline
 {data statistics} &   $\pm15.8$ & $\pm5.9$ & $\pm3.4$ & $\pm3.1$ &  $\pm4.7$  &  $\pm9.0$ \\
 \hline
  \hline
\multicolumn{7}{|c|}{$\sigma(\dsmucross X)$ and $\sigma(\Bcross)$ relative systematic error (\%)} \\
  \hline
 $ {\dstaro\mu}$ fit &   $\pm 3.5$    & $\pm1.8$   & $\pm 1.0$   &  $\pm 1.4$    & $\pm 1.7$   & $\pm 2.0$\\     
 \hline
  $f_b$               &   $^{+2.5}_{-3.8}$    & $^{+2.3}_{-3.5}$   & $^{+1.8}_{-2.8}$   &  $^{+1.6}_{-2.5}$    & $^{+1.4}_{-2.2}$   & $^{+1.8}_{-2.9}$\\            
 \hline
 ${\mu}$ trigger &   $^{+1.3}_{-1.2}$ & $^{+1.3}_{-1.3}$& $^{+1.7}_{-1.6}$&  $^{+2.2}_{-2.0}$    & $^{+2.5}_{-2.2}$   & $^{+2.7}_{-2.5}$\\   
 \hline
 tracking + $\mu$ reconstruction  &   $^{+9.1}_{-8.2}$    & $^{+9.0}_{-8.1}$   & $^{+8.9}_{-8.0}$   &  $^{+8.7}_{-7.9}$    & $^{+8.5}_{-7.7}$   & $^{+8.3}_{-7.5}$\\            
 \hline
 MC $\pT/\eta$ reweight     &       $^{+0.2}_{-1.3}$ & $^{+0.2}_{-1.2}$& $^{+0.4}_{-1.1}$&  $^{+0.5}_{-1.1}$    & $^{+0.4}_{-1.0}$   & $^{+0.2}_{-0.8}$\\   
  \hline
 $D^0$ and $H_b$ vertices fit     &       $\pm2.0$ & $\pm2.0$& $\pm 2.0$&  $\pm2.0$    & $\pm2.0$   & $\pm2.0$\\   
  \hline
 $D^0$ mass correction           &     $^{+0.8}_{-1.0}$ & $^{+0.8}_{-1.0}$& $^{+0.8}_{-1.0}$&  $^{+0.8}_{-1.0}$    & $^{+0.8}_{-1.0}$   & $^{+0.8}_{-1.0}$\\   
 \hline
 luminosity          & \multicolumn{6}{c|}{$\pm 3.4$} \\
  \hline
  $\mathscr{B}({\dstarp\rightarrow D^0\pi^+})$   & \multicolumn{6}{c|}{$\pm 0.7$} \\ 
 \hline
  $\mathscr{B}({D^0 \rightarrow K^- \pi^+})$   & \multicolumn{6}{c|}{$\pm 1.3$} \\ 
 \hline
  \hline
   \multicolumn{7}{|c|}{$\sigma(\dsmucross X)$ relative systematic error (\%)} \\
  \hline
 unfolding           &     $^{+6.6}_{-10.0}$ & $^{+2.3}_{-3.8}$& $^{+1.7}_{-1.5}$&  $^{+2.3}_{-1.3}$    & $^{+3.2}_{-6.7}$   & $^{+9.1}_{-3.5}$\\   
  \hline
  \hline
  
\multicolumn{7}{|c|}{$\sigma(\Bcross)$ relative systematic error in (\%)} \\
  \hline
 $\BR$   & \multicolumn{6}{c|}{$\pm 7$} \\  
\hline
$\mbox{unfolding $\oplus$ acceptance }$                    &   $^{+3.4}_{-11.3}$    & $^{+1.6}_{-6.4}$   & $^{+2.3}_{-4.0}$   &  $^{+0.7}_{-2.5}$    & $^{+1.7}_{-4.4}$   & $^{+6.0}_{-1.0}$\\ 
\hline
  \hline
 total syst $\sigma(\dsmucross X)$ &   $^{+12.9}_{-14.7}$    & $^{+10.7}_{-10.8}$   & $^{+10.3}_{-9.9}$   &  $^{+10.4}_{-9.8}$    & $^{+10.6}_{-11.7}$   & $^{+13.6}_{-10.3}$\\     
\hline
  total syst $\sigma(\Bcross)$ &   $^{+13.4}_{-17.1}$    & $^{+12.6}_{-14.0}$   & $^{+12.5}_{-12.5}$   &  $^{+12.3}_{-12.1}$    & $^{+12.3}_{-12.5}$   & $^{+13.5}_{-11.8}$\\     
\hline
\end{tabular}
\end{center}
\caption{$\dsmucross X$ and $\Bcross$ cross section relative uncertainties as a function of $\pT(\Bhadrn)$, listed as percentages (\%).}
\label{syst1}
\end{table}

%% file: tab_08.tex
\begin{table}[!h]
\renewcommand{\arraystretch}{1.1}
\begin{center}
\begin{tabular}{|l||c|c|c|c|c||c|}
\hline
 $|\eta|$ bin                        &       0--0.5      & 0.5--1     & 1--1.5      &  1.5--2    &  2--2.5  & 0--2.5     \\
 \hline
 \hline
 {data statistics} &   $\pm3.9$ & $\pm4.3$ & $\pm5.8$ & $\pm7.3$ &  $\pm12.5$ & $\pm2.5$ \\
 \hline
 \hline
\multicolumn{7}{|c|}{$\sigma(\dsmucross X)$ and $\sigma(\Bcross)$ relative systematic error (\%)} \\
  \hline
 ${\dstaro\mu}$ fit &   $\pm 0.7$    & $\pm 0.9$   & $\pm 0.7$   &  $\pm 1.2$    & $\pm 1.0$  & $\pm 0.5$\\     
 \hline
  $f_b$               &   $^{+1.6}_{-2.6}$    & $^{+2.0}_{-3.5}$   & $^{+1.5}_{-2.4}$   &  $^{+1.5}_{-2.6}$    & $^{+1.3}_{-2.1}$ & $^{+1.7}_{-2.8}$ \\            
 \hline
 $\mu$ trigger &   $^{+2.0}_{-1.9}$ & $^{+2.1}_{-1.9}$& $^{+1.8}_{-1.6}$&  $^{+1.7}_{-1.6}$    & $^{+1.6}_{-1.5}$  & $^{+1.9}_{-1.9}$\\   
 \hline
 tracking + $\mu$ reconstruction  &   $^{+7.0}_{-6.5}$    & $^{+7.1}_{-6.6}$   & $^{+8.5}_{-7.7}$   &  $^{+11.4}_{-10.0}$    & $^{+16.2}_{-13.4}$  & $^{+8.6}_{-8.0}$ \\            
 \hline
 MC $\pT/\eta$ reweight  &    $^{+1.5}_{-0.1}$ & $^{+1.2}_{-0.1}$& $^{+1.4}_{-0.1}$&  $^{+1.1}_{-0.2}$    & $^{+2.0}_{-0.5}$ & $^{+1.3}_{-1.3}$ \\      
  \hline
 $D^0$ and $H_b$ vertices fit     &       $\pm2.0$ & $\pm2.0$& $\pm 2.0$&  $\pm2.0$    & $\pm2.0$ & $\pm2.0$ \\   
           \hline
 $D^0$ mass correction   &    $^{+0.8}_{-1.0}$ & $^{+0.8}_{-1.0}$& $^{+0.8}_{-1.0}$&  $^{+0.8}_{-1.0}$    & $^{+0.8}_{-1.0}$& $^{+0.8}_{-1.0}$   \\                    
   \hline
 luminosity         & \multicolumn{6}{c|}{$\pm3.4$} \\ 
 \hline
  $\mathscr{B}({\dstarp\rightarrow D^0\pi^+})$   & \multicolumn{6}{c|}{$\pm0.7$} \\ 
  \hline
  $\mathscr{B}({D^0 \rightarrow K^- \pi^+})$     & \multicolumn{6}{c|}{$\pm1.3$} \\ 
   \hline
 \hline
\multicolumn{7}{|c|}{$\sigma(\dsmucross X)$ relative systematic error (\%)} \\
   \hline
 unfolding  &    $^{+1.3}_{-0.9}$ & $^{+1.1}_{-1.5}$& $^{+1.4}_{-0.8}$&  $^{+0.7}_{-1.0}$    & $^{+1.1}_{-2.0}$ & - \\                  
 \hline
 \hline
\multicolumn{7}{|c|}{$\sigma(\Bcross)$ relative systematic error (\%)} \\
  \hline
$\BR$  & \multicolumn{6}{c|}{$\pm7$} \\ 
\hline
$\mbox{unfolding $\oplus$ acceptance}$                    &   $^{+5.1}_{-15.0}$    & $^{+7.3}_{-16.2}$   & $^{+10.7}_{-19.1}$   &  $^{+4.8}_{-24.6}$    & $^{+4.3}_{-29.6}$ & $^{+6.4}_{-17.1}$ \\ 
\hline
   \hline
 total syst $\sigma(\dsmucross X)$ &   $^{+8.8}_{-8.5}$    & $^{+9.0}_{-9.0}$   & $^{+10.0}_{-9.4}$   &  $^{+12.5}_{-11.4}$    & $^{+17.1}_{-14.5}$ & $^{+10.0}_{-9.8}$ \\     
\hline
 total syst $\sigma(\Bcross)$ &   $^{+12.2}_{-18.5}$    & $^{+13.4}_{-19.7}$   & $^{+16.1}_{-22.3}$   &  $^{+15.0}_{-27.9}$    & $^{+18.8}_{-33.5}$ &  $^{+13.8}_{-20.9}$\\     
\hline

\end{tabular}
\end{center}
\caption{$\dsmucross X$ and $\Bcross$ cross section relative uncertainties as a function of $|\eta(\Bhadrn)|$, listed as percentages (\%). The last column refers to the integrated cross sections.}
\label{syst2}
\end{table}

%% file: sigma.tex
\section{Differential cross sections for $b$-hadron production}
\label{sec:acc}

The $b$-hadron differential cross sections can be derived from the $\dsmucross X$ differential cross sections by taking into account the branching ratio $\BR$ and the necessary decay acceptance corrections. 
These are evaluated using a {\sc Powheg+Pythia} simulation in two steps:
\begin{itemize}
	\item Identification of the $\Bhadrn$ kinematic region selected by the $\dstarp$ and $\mu^-$ kinematic cuts. This indicates that only $b$-hadrons with $\pT(\Bhadrn)>9\GeV$ and $|\eta(\Bhadrn)|<2.5$ pass the $\dstarp$ and $\mu^-$ kinematic cuts. 
	\item Evaluation of a bin-by-bin $\pT$- and $|\eta|$-decay acceptance $\alpha$ in the $\Bhadrn$ allowed kinematic region, defined as
	\begin{equation}
		\alpha = \frac{\mbox{number of }\Bhadrn(\rightarrow \dsmu) \mbox{ passing the }\dstaro\mbox{ and }\mu \mbox{ kinematic cuts}}{\mbox{number of }\Bhadrn(\rightarrow \dsmu) \mbox{ passing the }\Bhadrn \mbox{ kinematic cuts}}
	\end{equation}
	
\end{itemize}
The results are shown in Table \ref{tab:alpha} for the {\sc Powheg+Pythia} central prediction. Section \ref{sec:syst} describes how the NLO theoretical uncertainties are propagated to this measurement. 

\input{tab_09}

The $b$-hadron differential cross sections as a function of $\pT$ and $\eta$, inside the kinematic region $\pT(\Bhadrn)>9\GeV$ and $|\eta(\Bhadrn)|<2.5$, can then be calculated according to the formula:
\begin{equation}
      \frac{\mathrm{d}\sigma(\Bcross \,X)}{\mathrm{d}\pT(\eta)} = \frac{1}{\alpha_{\pT (\eta)} \BR }\frac{\mathrm{d}\sigma(pp\to\dsmucrossext )}{\mathrm{d}\pT(\eta)}
  \end{equation}

\begin{figure}[!h]
  \subfigure[]{
  \includegraphics[angle=0,width=0.495\columnwidth]{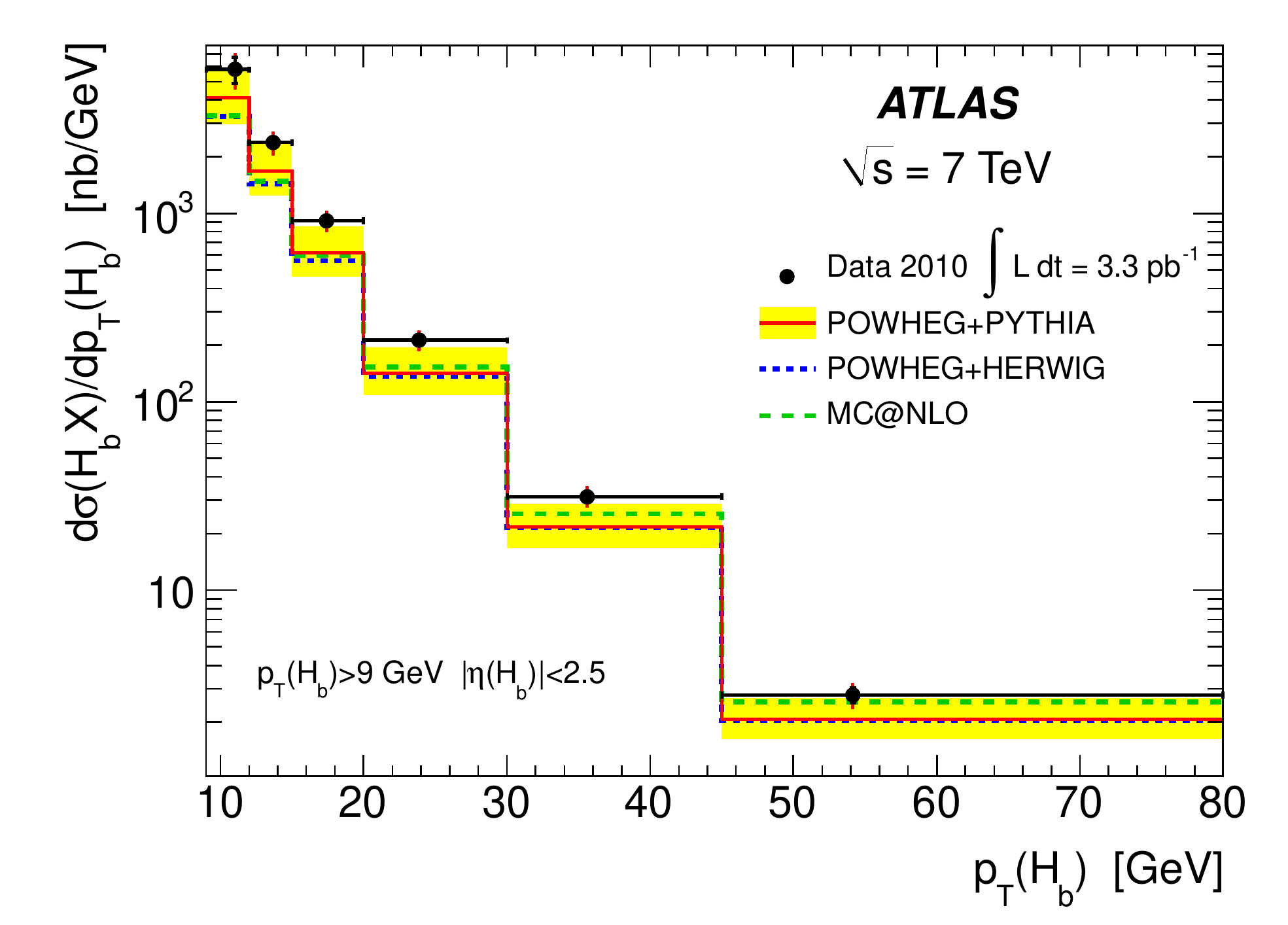}
  }
    \subfigure[]{
     \includegraphics[angle=0,width=0.495\columnwidth]{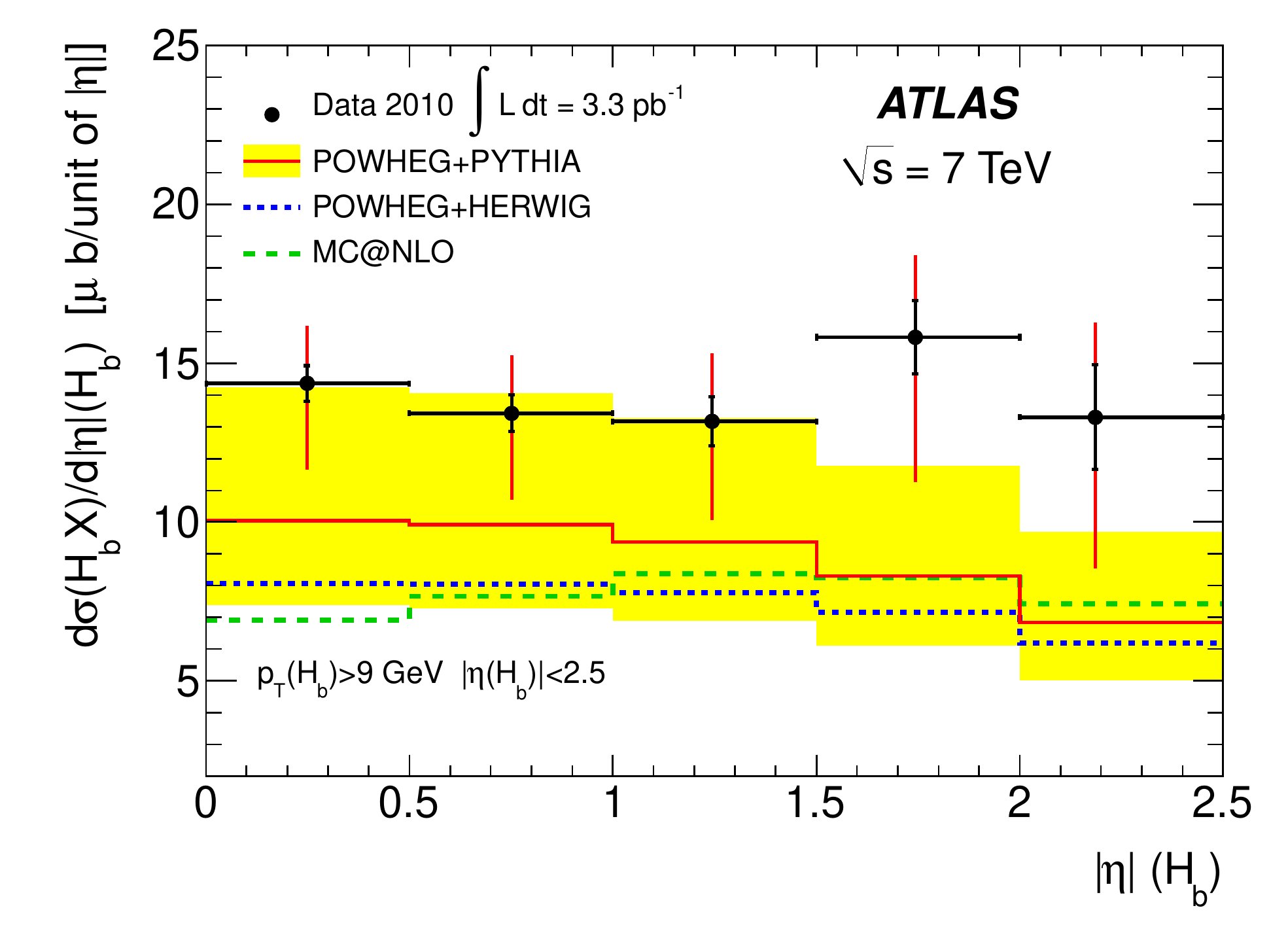}
  }

    \caption{Differential cross section for $\Bhadrn$ production as a function of (a) $\pT$ and (b) $|\eta|$ of the $b$-hadron, in the fiducial kinematical region $\pT(\Bhadrn)>9\GeV, |\eta(\Bhadrn)|<2.5$. The measurement is compared with the theoretical predictions, as described in the text. The inner error bars of the data points are statistical uncertainties, the outer are statistical+total systematic uncertainties. 
    }

\label{Btotcross}
  \end{figure}

Figure~\ref{Btotcross} shows  the $b$-hadron differential cross section measurements compared with theoretical predictions. The shaded band is the overall theoretical uncertainty of the central {\sc Powheg+Pythia} prediction. Since the acceptance correction factors have a dependence on $\pT$ and $|\eta|$, as shown in Table~\ref{tab:alpha}, the shapes of the $b$-hadron differential cross sections are different to the $\dsmucross X$ differential cross sections shown in Figure~\ref{crossbdsmu}. The systematic uncertainties are those from the $\sigma(\dsmucross X)$ measurement described in Section~\ref{sec:unfold}, with the addition of the uncertainty of the branching ratio $\BR$ and the uncertainties of the decay acceptance correction. The $b$-hadron differential cross section values are reported in Table \ref{tab:Xsections}, together with the statistical and total systematic uncertainties, while the individual contributions to the systematic uncertainty are reported in Tables \ref{syst1} and \ref{syst2}. The combined unfolding and acceptance uncertainties are calculated taking their correlations into account.

The comparison with data shows that NLO calculations underestimate the cross section, although the difference is within the combined experimental and theoretical uncertainties.
The $b$-hadron integrated cross section for $\pT(\Bhadrn)>9\GeV$ and $|\eta (\Bhadrn)|<2.5$ is measured as:
$$\sigma(pp\rightarrow\Bcross \,X) = 32.7 \pm 0.8\stat \pm 3.1\syst ^{+2.1}_{-5.6}(\alpha)\pm 2.3(\mathscr{B}) \pm 1.1(\mathscr{L})  \,\,\mu\mbox{b}$$

The integrated {\sc Powheg+Pythia} prediction, with its theoretical uncertainty, is:

$$\sigma(pp\rightarrow\Bcross X) = 22.2 \, ^{+8.9}_{-5.4} (\mbox{scale}) \, ^{+2.1}_{-1.9} (m_b) \, ^{+2.2}_{-2.1} (\mbox{PDF}) \, ^{+1.6}_{-1.5} (\mbox{hadr.}) \,\,\mu\mbox{b}$$

The corresponding {\sc Powheg+Herwig} prediction is 18.6\,$\mu$b, while {\sc mc@nlo} predicts 19.2\,$\mu$b, with similar theoretical uncertainties to the {\sc Powheg+Pythia} prediction.

%% file: tab_09.tex
\begin{table}[h]
\centering
\subtable{
\begin{tabular}{|c|c|}
\hline
  $\pT(\Bhadrn)$  & $\alpha$  \\
 \hline
 \hline
 $\phantom{0}$9--12 \GeV &   $0.005$  \\
 \hline
 12--15 \GeV               &   $0.071$  \\ 
 \hline
  15--20 \GeV &   $0.219$  \\
 \hline
 20--30 \GeV               &   $0.422$  \\ 
 \hline
 30--45 \GeV &   $0.614$  \\
 \hline
 45--80 \GeV               &   $0.723$ \\ 
 \hline
\end{tabular}
}\qquad\qquad
\subtable{
\begin{tabular}{|c|c|}
\hline
  $|\eta(\Bhadrn)|$   &  $\alpha$  \\
 \hline
 \hline
 0.0--0.5 &   $0.096$  \\
 \hline
 0.5--1.0               &   $0.095$  \\ 
 \hline
  1.0--1.5&   $0.091$  \\
 \hline
 1.5--2.0               &   $0.086$  \\ 
 \hline
 2.0--2.5 &   $0.061$  \\
 \hline
\end{tabular}
}
\caption{Decay acceptance $\alpha$ as a function of $\pT(\Bhadrn)$ and $|\eta(\Bhadrn)|$ for the {\sc Powheg+Pythia} prediction.}
\label{tab:alpha}
\end{table}

%% file: discussion.tex
\section{Discussion}
\label{sec:discussion}

Section~\ref{sec:acc} discusses the measurement of the $b$-hadron production cross section for $\pT(\Bhadrn)>9\GeV$ and $|\eta(\Bhadrn)|<2.5$.
In order to  compare this result with other LHC measurements, we extrapolate this measurement to the full kinematic phase space, extending to regions outside the ATLAS coverage, using the NLO MC theoretical predictions.
The multiplicative extrapolation factor is defined as the ratio of the total number of generated $b$-hadrons to the number of $b$-hadrons generated with  $\pT(\Bhadrn)>9\GeV$ and $|\eta(\Bhadrn)|<2.5$, and is estimated to be $11.0 ^{+2.6}_{-1.6}$. The resulting total $b$-hadron cross section is:
$$
\sigma(pp\rightarrow \Bcross \,X)_\mathrm{total} = 360 \pm 9\stat \pm 34\syst \pm 25(\mathscr{B}) \pm 12(\mathscr{L}) ^{+77}_{-69} (\mbox{accept.}\oplus\mbox{extrap.}) \,\,\mu\mbox{b}
$$
where the combined acceptance and extrapolation uncertainty is calculated taking their correlations into account.

This value can be compared with the inclusive $\bbbar$ cross section measurements by LHCb $\sigma(pp \rightarrow b\bar b X) = 284 \pm 20 \stat \pm 49 \syst \,\, \mu\mbox{b}$, evaluated in the kinematic region $2<\eta<6$ using decays to $D^0 \mu^- \overline{\nu}X$ final states~\cite{lhcb1}, and $\sigma(pp \rightarrow b\bar b X) = 288 \pm 4 \stat \pm 48 \syst \,\, \mu\mbox{b}$, evaluated using $J/\psi X$ final states in the kinematic region $2.0<y<4.5$~\cite{lhcb2}. Extrapolations outside the LHCb sensitivity region are done using different theoretical models, without including additional uncertainties. Also ALICE measured the inclusive $\bbbar$ cross section in $pp$ collisions, using decays to $J/\psi X$ final states in the kinematic region $|y|<0.9$ and $p_T>1.3\GeV$~\cite{alice}. After extrapolation to the full phase space, they obtain $\sigma(pp \rightarrow b\bar b X) = 244 \pm 64 \stat ^{+50}_{-59} \syst ^{+7}_{-6} \mbox{(extr.)}    \,\, \mu\mbox{b}$.

%% file: conclusion.tex
\section{Conclusions}
\label{sec:conclusions}

The production of $b$-hadrons ($\Bcross$) at the LHC is measured with the ATLAS detector in proton-proton collisions at $\sqrt{s}=7\TeV$, using 3.3\,\ipb{} of integrated luminosity from the 2010 run. A $b$-hadron enriched sample was obtained by combining oppositely charged $\Dstar$ mesons and muons, in events triggered by a muon with $\pT$ exceeding 6\,\GeV.

Differential cross sections as functions of $\pT$ and $|\eta|$ are produced for both $\Bcross$ and $\dsmucross X$ production. These measurements are found to be higher than the NLO QCD predictions, but consistent within the experimental and theoretical uncertainties. The integrated $b$-hadron cross section for $\pT(\Bhadrn)>9\GeV$ and $|\eta(\Bhadrn)|<2.5$ is measured as 
\begin{linenomath}
$$\sigma(pp\rightarrow\Bcross \,X) = 32.7 \pm 0.8\stat \pm 3.1\syst ^{+2.1}_{-5.6}(\alpha)\pm 2.3(\mathscr{B}) \pm 1.1(\mathscr{L})  \,\,\mu\mbox{b}.$$
\end{linenomath}

%% file: acknowledgements.tex
\section{Acknowledgements}

We thank CERN for the very successful operation of the LHC, as well as the
support staff from our institutions without whom ATLAS could not be
operated efficiently.

We acknowledge the support of ANPCyT, Argentina; YerPhI, Armenia; ARC,
Australia; BMWF, Austria; ANAS, Azerbaijan; SSTC, Belarus; CNPq and FAPESP,
Brazil; NSERC, NRC and CFI, Canada; CERN; CONICYT, Chile; CAS, MOST and NSFC,
China; COLCIENCIAS, Colombia; MSMT CR, MPO CR and VSC CR, Czech Republic;
DNRF, DNSRC and Lundbeck Foundation, Denmark; EPLANET and ERC, European Union;
IN2P3-CNRS, CEA-DSM/IRFU, France; GNAS, Georgia; BMBF, DFG, HGF, MPG and AvH
Foundation, Germany; GSRT, Greece; ISF, MINERVA, GIF, DIP and Benoziyo Center,
Israel; INFN, Italy; MEXT and JSPS, Japan; CNRST, Morocco; FOM and NWO,
Netherlands; RCN, Norway; MNiSW, Poland; GRICES and FCT, Portugal; MERYS
(MECTS), Romania; MES of Russia and ROSATOM, Russian Federation; JINR; MSTD,
Serbia; MSSR, Slovakia; ARRS and MVZT, Slovenia; DST/NRF, South Africa;
MICINN, Spain; SRC and Wallenberg Foundation, Sweden; SER, SNSF and Cantons of
Bern and Geneva, Switzerland; NSC, Taiwan; TAEK, Turkey; STFC, the Royal
Society and Leverhulme Trust, United Kingdom; DOE and NSF, United States of
America.

The crucial computing support from all WLCG partners is acknowledged
gratefully, in particular from CERN and the ATLAS Tier-1 facilities at
TRIUMF (Canada), NDGF (Denmark, Norway, Sweden), CC-IN2P3 (France),
KIT/GridKA (Germany), INFN-CNAF (Italy), NL-T1 (Netherlands), PIC (Spain),
ASGC (Taiwan), RAL (UK) and BNL (USA) and in the Tier-2 facilities
worldwide.

%% file: atlas_authlist.tex
\begin{flushleft}
{\Large The ATLAS Collaboration}

\bigskip

G.~Aad$^{\rm 48}$,
B.~Abbott$^{\rm 111}$,
J.~Abdallah$^{\rm 11}$,
S.~Abdel~Khalek$^{\rm 115}$,
A.A.~Abdelalim$^{\rm 49}$,
O.~Abdinov$^{\rm 10}$,
B.~Abi$^{\rm 112}$,
M.~Abolins$^{\rm 88}$,
O.S.~AbouZeid$^{\rm 158}$,
H.~Abramowicz$^{\rm 153}$
% ...
\bigskip

% ...
$^{10}$ Institute of Physics, Azerbaijan Academy of Sciences, Baku, Azerbaijan\\
$^{11}$ Institut de F\'isica d'Altes Energies and Departament de F\'isica de la Universitat Aut\`onoma  de Barcelona and ICREA, Barcelona, Spain\\
% ...
$^{48}$ Fakult\"{a}t f\"{u}r Mathematik und Physik, Albert-Ludwigs-Universit\"{a}t, Freiburg i.Br., Germany\\
$^{49}$ Section de Physique, Universit\'e de Gen\`eve, Geneva, Switzerland\\
% ...
$^{88}$ Department of Physics and Astronomy, Michigan State University, East Lansing MI, United States of America\\
% ...
$^{111}$ Homer L. Dodge Department of Physics and Astronomy, University of Oklahoma, Norman OK, United States of America\\
$^{112}$ Department of Physics, Oklahoma State University, Stillwater OK, United States of America\\
% ...
$^{115}$ LAL, Universit\'e Paris-Sud and CNRS/IN2P3, Orsay, France\\
% ...
$^{153}$ Raymond and Beverly Sackler School of Physics and Astronomy, Tel Aviv University, Tel Aviv, Israel\\
% ...
$^{158}$ Department of Physics, University of Toronto, Toronto ON, Canada\\
% ...
\end{flushleft}


%% file: BXsec.bbl
\begin{thebibliography}{00}

\bibitem{theory1} P. Nason, S. Dawson and R.K. Ellis, ``The total cross section for the production of heavy quarks in hadronic collisions'', Nucl. Phys. B 303 (1988) 607.
\bibitem{theory2} P. Nason, S. Dawson and R.K. Ellis, ``The one particle inclusive differential cross section for heavy quark production in hadronic collisions'', Nucl. Phys. B 327 (1989) 49.
\bibitem{ua1} UA1 Collaboration, ``Beauty Production at the CERN Proton- anti-Proton Collider'', Phys. Lett. B 186 (1987) 237.
\bibitem{ua2} UA1 Collaboration, ``Measurement of the Bottom Quark Production Cross Section in Proton-anti-Proton Collisions at $\sqrt{s}=0.63\,\TeV{}$'', Phys. Lett. B 213 (1988) 405.

\bibitem{cdf0} CDF Collaboration, ``Measurement of the Ratio of the $b$ Quark Production Cross Sections in $p \bar p$ Collisions at $\sqrt{s}=630\,\GeV{}$ and $\sqrt{s}=1800\,\GeV{}$'', Phys. Rev. D 66 (2002) 032002.
\bibitem{cdf1} CDF Collaboration, ``Measurement of the bottom quark production cross-section using semileptonic decay electron in $p\bar p$ collisions at $\sqrt{s}=1.8\,\TeV{}$'', Phys. Rev. Lett. 71 (1993) 500.
\bibitem{cdf2} CDF Collaboration, ``Measurement of the $B$ meson differential cross-section, $d\sigma/d\pT$, in $p\bar p$ collisions at $\sqrt{s}=1.8\,\TeV{}$'', Phys. Rev. Lett. 75 (1995) 1451.
\bibitem{cdf3} CDF Collaboration, ``Measurement of the $B^+$ total cross section and $B^+$ differential cross section $d\sigma/d\pT$ in $p\bar p$ collisions at $\sqrt{s}=1.8\,\TeV{}$'', Phys. Rev. D 65 (2002) 052005.
\bibitem{d01} D0 Collaboration, ``Inclusive $\mu$ and $B$ quark production cross-sections in $p\bar p$ collisions at $\sqrt{s}=1.8\,\TeV{}$'', Phys. Rev. Lett. 74 (1995) 3548.
\bibitem{d02} D0 Collaboration, ``Small angle muon and bottom quark production in $p\bar p$ collisions at $\sqrt{s}=1.8\,\TeV{}$'', Phys. Rev. Lett. 84 (2000) 5478.
\bibitem{d03} D0 Collaboration, ``Cross section for $b$ jet production in $p\bar p$ collisions at $\sqrt{s}=1.8\,\TeV{}$'', Phys. Rev. Lett. 85 (2000) 5068.
\bibitem{cdf4} CDF Collaboration, ``Measurement of the $\jpsi$ meson and $b$-hadron production cross sections in $p\bar p$ collisions at $\sqrt{s}=1960\,\GeV{}$'', Phys. Rev. D 71 (2005) 032001.
\bibitem{cdf5} CDF Collaboration, ``Measurement of the $B^+$ production cross section in $p\bar p$ collisions at $\sqrt{s}=1960\,\GeV{}$'', Phys. Rev. D 75 (2007) 012010.
\bibitem{cdf6} CDF Collaboration, ``Measurement of the $b$-hadron production cross section using decays to $\mu^-D^0 X$ final states in $p\bar p$ collisions at $\sqrt{s}=1960\,\GeV{}$'', Phys. Rev. D 79 (2009) 092003.

\bibitem{cacciari} M. Cacciari, S. Frixione, M. Mangano et al., ``QCD analysis of first b cross-section data at 1.96\TeV{}'', JHEP 0407 (2004) 033.

\bibitem{lhcb1} LHCb Collaboration, ``Measurement of $\sigma({b\bar b})$ at $\sqrt{s}=7\TeV$ in the forward region'', Phys. Lett. B 694 (2010) 209.
\bibitem{lhcb2} LHCb Collaboration, ``Measurement of $J/\psi$ production in $pp$ collisions at $\sqrt{s}=7\TeV$'', Eur. Phys. J. C 71 (2011) 1645.
\bibitem{lhcb3} LHCb Collaboration, ``Measurement of the $B^{\pm}$ production cross-section in $pp$ collisions at $\sqrt{s}=7\TeV$'', arXiv:1202.4812.


\bibitem{cms1} CMS Collaboration, ``Measurement of the $B^+$ Production Cross Section in $pp$ Collisions at $\sqrt{s}=7\TeV$'', Phys. Rev. Lett. 106 (2011) 112001.
\bibitem{cms2} CMS Collaboration, ``Measurement of the $B^0$ Production Cross Section in $pp$ Collisions at $\sqrt{s}=7\TeV$'', Phys. Rev. Lett. 106 (2011) 252001.
\bibitem{cms3} CMS Collaboration, ``Measurement of the $B_s^0$ Production Cross Section with $B_s^0\to J/\Psi \phi$ Decays in $pp$ Collisions at $\sqrt{s}=7\TeV$'', Phys. Rev. D 84 (2011) 052008.
\bibitem{cms4} CMS Collaboration, ``Inclusive $b$-hadron production cross section with muons in $pp$ collisions at $\sqrt{s}=7\TeV$'', JHEP 03 (2011) 090.
\bibitem{cms5} CMS Collaboration, ``Measurement of the cross section for production of $b \bar b X$, decaying to muons in $pp$ collisions at $\sqrt{s}=7\TeV$'', arXiv:1203.3458.

\bibitem{alice} ALICE Collaboration, ``Measurement of prompt and non-prompt $J/\psi$ production cross sections at mid-rapidity in $pp$ collisions at $\sqrt{s}=7\TeV$'', arXiv:1205.5880v1.


\bibitem{atlas} ATLAS Collaboration, ``The ATLAS Experiment at the CERN Large Hadron Collider'', JINST 3 S08003 (2008). 

\bibitem{PDG} K. Nakamura et al. (Particle Data Group), J. Phys. G 37, 075021 (2010). 
\bibitem{pythia} T. Sjoestrand, S. Mrenna, P. Skands, ``{\sc Pythia} 6.4 Physics and Manual'', JHEP 05 (2006) 026. 
\bibitem{atlastune} ATLAS Collaboration, ``Charged-particle multiplicities in pp interactions measured with the ATLAS detector at the LHC'', New J. Phys. 13 (2011) 053033. 
\bibitem{geant} ATLAS Collaboration, ``The ATLAS Simulation Infrastructure'', Eur. Phys. J. C 70 (2010) 823.
\bibitem{geant2} S. Agostinelli et al., ``GEANT4: A simulation toolkit'', Nucl. Instrum. Meth. A 506 (2003) 250. 
\bibitem{mcnlo1} S. Frixione and B. R. Webber, ``Matching NLO QCD computations and parton shower simulations'', JHEP 0206 (2002) 029.
\bibitem{mcnlo2} S. Frixione, P. Nason and B. R. Webber, ``Matching NLO QCD and parton showers in heavy flavour production'', JHEP 0308 (2003) 007.
\bibitem{pow1} P. Nason, ``A new method for combining NLO QCD with shower Monte Carlo algorithms'', JHEP 0411 (2004) 040. 
\bibitem{pow2} S. Frixione, P. Nason and G. Ridolfi, ``A Positive-Weight Next-to-Leading-Order Monte Carlo for Heavy Flavour Hadroproduction'', JHEP 0709 (2007) 126.
\bibitem{herwig1} G. Corcella et al. , ``{\sc Herwig} 6: an event generator for Hadron Emission Reactions With Interfering Gluons'', JHEP 0101 (2001) 010.
\bibitem{cteq} P. M. Nadolsky, ``Implications of CTEQ global analysis for collider observables'', Phys. Rev. D 78 (2008) 013004.
\bibitem{cluster} B. R. Webber, ``A QCD model for jet fragmentation including soft gluon interference'', Nucl. Phys. B 238 (1984) 492.
\bibitem{lund} B. Anderson et al., ``Parton fragmentation and string dynamics'', Phys. Rep. 97 (1983) 31.
\bibitem{bowler} M. G. Bowler, ``$e^+e^-$ Production of Heavy Quarks in the String Model'', Z. Phys. C 11 (1981) 169.
\bibitem{lund2} B. Anderson, G. Gustafson, B. Soederberg, ``A General Model for Jet Fragmentation'', Z. Phys. C 20 (1983) 317.
\bibitem{peterson} C. Peterson et al., ``Scaling violations in Inclusive $e^+e^-$ Annihilation Spectra'', Phys. Rev. D 27 (1983) 105.


\bibitem{jpsi} ATLAS Collaboration, ``Measurement of the differential cross-sections of inclusive, prompt and non-prompt $J/\psi$ production in proton-proton collisions at $\sqrt{s}=7\TeV$", Nucl. Phys. B 850 (2011) 387. 

\bibitem{lumi1} ATLAS Collaboration, ``Luminosity Determination  in pp Collisions at $\sqrt{s}=7\TeV$ Using the ATLAS Detector at the LHC'', Eur. Phys. J. C 71 (2011) 1630.
\bibitem{lumi2} ATLAS Collaboration, ``Updated Luminosity Determination in pp Collisions at $\sqrt{s}=7$ TeV using the ATLAS Detector'', ATLAS-CONF-2011-011, 2011.

\bibitem{unfold1} G. Cowan, ``A Survey of Unfolding Methods for Particle Physics'', Proc. Advanced Statistical Techniques in Particle Physics, Durham (2002).
\bibitem{unfold2} G. Cowan, ``Statistical Data Analysis'', Oxford University Press (1998).
\bibitem{unfold3} V. Blobel, ``Unfolding Methods in High Energy Physics'', DESY 84-118 (1982), CERN 85-02 (1985).
\bibitem{unfold4} G. Zech, ``Comparing statistical data to Monte Carlo simulation - parameter fitting and unfolding'', DESY 95-113 (1995).
\bibitem{bayesunf} G. D'Agostini, ``A Multidimensional unfolding method based on Bayes' theorem'', Nucl. Instrum. Meth. A 362 (1995) 487.
%
\ifthenelse{\boolean{supp}}{
\bibitem{vadim} V. Kostioukhine, ``VKalVrt - package for vertex reconstruction in ATLAS'', ATLAS-PHYS-2003-031, 2003.
\bibitem{roounfold} The RooUnfold package and documentation are available from:\\ { http://hepunx.rl.ac.uk/$\sim$adye/software/unfold/RooUnfold.html}.
\bibitem{bjet} ATLAS Collaboration, ``Measurement of the inclusive and dijet cross-sections of $b$-jets in $pp$ collisions at $\sqrt{s}=7\,\TeV$ with the ATLAS detector'', submitted to EPJC (30 Sept 2011), arXiv:1109.6833v1 [hep-ex].
}{}


 \end{thebibliography}
